\documentclass[]{aa}
\usepackage{epsfig}

\begin{document}

\title{OH spectral evolution of oxygen-rich late-type stars}
\date{version 1.0\\ \today}
\author{S. Etoka \inst{1} 
	\and A.M. Le Squeren \inst{2}}
\institute{  Jodrell Bank Observatory, University of Manchester, 
	     Macclesfield, Cheshire SK11 9DL, UK
        \and GEPI, Observatoire de Paris-Meudon, 5 place J. Janssen, 
	     F-92195 Meudon Cedex, France}

\offprints{ S. Etoka}
\mail{setoka@jb.man.ac.uk}
\date{Received date; accepted 27 January 2004}
%
\abstract{
We investigated the main-line spectral evolution with shell thickness of 
oxygen rich AGB stars. The study is based on a sample of 30 sources 
distributed along the IRAS colour-colour diagram. The sources were chosen to 
trace the Miras with thick shells and the whole range of OH/IR stars.
The Miras exhibit a 1665~MHz emission strength comparable to that 
at 1667~MHz. Even though the Miras of the study have quite thick shells, 
their spectral characteristics in both main lines attest to a strong 
heterogeneity in their OH shell with, in particular, the presence of 
significant turbulence and acceleration. The expansion velocity has been 
found to be about the same at 1665 and 1667~MHz, taking into account a 
possible velocity turbulence of 1-2~km~s$^{-1}$ at the location of the 
main-line maser emission. An increase in the intensity ratio 1667/1665 with 
shell thickness has been found. A plausible explanation for such a 
phenomenon is that competitive gain in favour of the 1667~MHz line increases 
when the shell is getting thicker. 
There is an evolution in the spectral profile shape with the appearance of a 
substantial inter-peak signal when the shell is getting thicker. Also, 
inter-peak components are found and can be as strong as the 
external standard peaks when the shell is very thick. This trend for an 
increase of the signal in between the two main peaks is thought 
to be the result of an increase of the saturation with shell thickness. 
All sources but two - a Mira and an OH/IR star from the lower 
part of the colour$-$colour diagram - are weakly polarized. The strong 
polarization observed for those two particular objects is thought to be 
the result of perturbations in their shells.
\keywords{Stars: AGB and post-AGB $-$ circumstellar matter $-$ Masers $-$
          Line: profiles $-$ Stars: mass-loss $-$ Stars: evolution}
}
\maketitle
\section{Introduction} \label{section Intro.}

The standard model proposed by Reid et~al. (1977) explains convincingly 
the double-peaked profile usually observed for Miras and OH/IR objects, 
particularly at 1612~MHz. It also explains the general features of the 
circumstellar envelopes. Nevertheless, many stars do not have such a simple 
OH maser profile. In particular, in the main lines, the peaks are often 
broad and consist of many components. To explain these more complex 
profiles new models have been suggested.

	Alcock \& Ross (1986) have shown that it is possible to change the 
profile shape for a partially or fully saturated maser when the masering 
region becomes sufficiently thick. When that happens, the 
profiles consist of broader peaks and a signal is seen at the velocity of the 
star. This is in agreement with the spectral profiles observed for Miras and 
OH/IR objects, but the inferred 
internal and external radii disagree with the MERLIN and VLA 
interferometric observations. To overcome this discrepancy, Alcock \& 
Ross suggest that the mass loss is not a continuous and uniform process 
but rather consists of ejection of blobs of material in random directions 
such that the symmetrical geometry is preserved. The model calculation 
results of Collison \& Nedoluha (1995) agree with the conclusions by 
Alcock \& Ross 
that the observed OH profile cannot be produced by the standard model 
description of a smooth, spherically symmetric, steady windlike mass loss. 
The assumption of a non-smooth shell is justified by the interferometric maps 
themself. Indeed, with the increase in interferometer resolution the more 
detailed maps revealed that the OH maser emission is located in clumps 
(Welty, Fix \& Mutel, 1987; Chapman, Cohen \& Saikia, 1991)
and/or the shell shows evidence for deviation from spherical symmetry 
(Diamond et~al. 1985; Bowers, Johnston \& de~Vegt 1989). 

Evidence for acceleration in the OH shell has been found for many stars  
(Chapman et~al. 1994; Etoka \& Le~Squeren 1996, 1997; 
 Szymczak et~al. 1998; Richards et~al. 1999).
The importance of acceleration in the shape of the spectral profile is 
underlined by the work of Chapman \& Cohen (1985). According to the value of 
the logarithmic velocity gradient $\epsilon$, the shape of the observed line 
profile largely varies as follows~:
(1) for $\epsilon \rightarrow 0$ the classical double-peaked profile is 
observed;
(2) for $0 < \epsilon < 1$ an inter-peak signal can be seen; 
(3) for $\epsilon \simeq 1$ a plateau-shaped profile is obtained and finally 
(4) for $\epsilon > 1$
the profile is reduced to a single peak centred on the stellar velocity. Thus, 
many spectral profiles resembling those observed can be modelled by 
introducing acceleration in the OH shell. 

Based on the IRAS measurements at 12, 25 and 60~$\mu$m, Olnon et~al. 
(1984) discovered the existence of a continuous sequence from Miras to OH/IR 
objects in the so-called colour-colour diagram which plots the colour indices 
$[60-25]$ vs. 
$[12-25]$. This sequence has been analysed in terms of shell thickness 
by van~der~Veen \& Habing (1988) and van~der~Veen \& Rugers (1989). 
These studies lead to a commonly accepted interpretation that it is an 
evolutionary track followed by intermediate-mass stars. In this scenario, 
Miras are the progenitors of OH/IR stars involving among other characteristics 
an increase of their mass-loss rate and expansion velocity with time. 
Nonetheless, 
the initial mass of the star could play a role in the setting up of that 
sequence. Based on an alternative model for the evolution of AGB stars 
proposed by Epchtein et~al. (1990), L\'epine et~al. (1995) 
proposed that the IRAS colour-colour sequence could be a sequence of mass. 

The study presented here deals with the spectral profiles in both main lines 
of Miras and OH/IR stars distributed along the colour-colour diagram. The main 
aim is to determine whether there is a relation between the spectral profile 
and the thickness of the circumstellar shell.  The next section presents the 
observations and selection criteria. Section~\ref{section Results} and 
Appendices~\ref{appendix A},~\ref{appendix B} and \ref{appendix C} present 
the results concerning the individual stars and a comparison with previous 
observations. The discussion is presented in Sect.~\ref{section Discus.} 
and the final conclusions in Sect.~\ref{section Conclu.}.

\section{Observations}  \label{section Observ.}

The OH observations were carried out between March and December 1994 with 
the {Nan\c cay} transit radiotelescope. The half-power beamwidth of the 
radiotelescope at 1.6~GHz is 3.5$\arcmin$ in $\alpha$ by 19$\arcmin$ in 
$\delta$. The system temperature was about 50 K. A frequency switching mode 
was used. 
The autocorrelator was divided into 4 banks of 256 channels each, allowing 
simultaneous observations of the two OH main lines (i.e., 1665 and 1667\,MHz) 
in both left- and right-handed circular polarizations (LHC, RHC). A bandwidth 
of 1.56~MHz was used, leading to a velocity resolution of 0.56~km~s$^{-1}$.
The ratio of flux to antenna temperature was 1.1~${\rm~Jy~K^{-1}}$ at 
${\rm 0^{o}}$ declination. Each observation lasted about one hour providing 
a mean rms of 0.035~K. All the radial velocities given hereafter are relative 
to the Local Standard of Rest (LSR).

\subsection{Selection criteria}

\begin{figure}[ht]
  \begin{center}
\hspace*{-0.5cm}        \epsfig{file=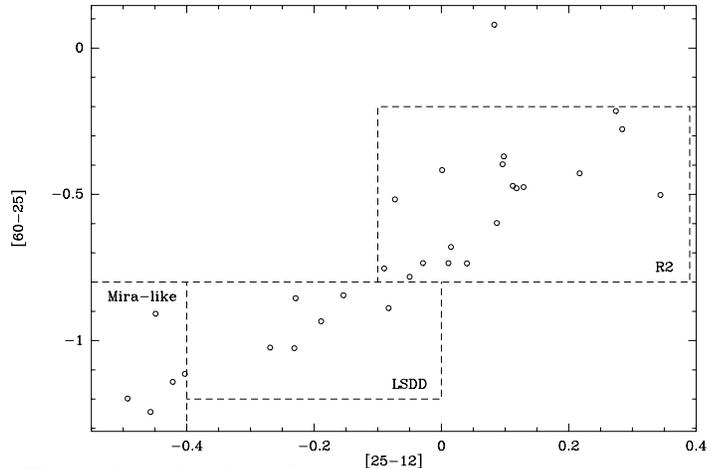,angle=-90,width=10cm}
  \end{center}
\vspace{-1.0cm}
\caption[]{\small [60-25] vs. [25-12] colour-colour diagram of the sources 
		  of the study}
\label{diagcoul}
\end{figure}

All the sources presented here had already been positively detected at 
1667~MHz and a few of them at 1665~MHz too. We looked for 1665~MHz emission 
for a possible signature of circumstellar shell evolution.
Figure~\ref{diagcoul} shows the colour-colour diagram of the selected sources. 
The colour indices [25$-$12] and [60$-$25] used here were calculated 
according to the following relation~:

\begin{equation}
   \label{def. couleur}
	 [\lambda_{1}-\lambda_{2}] = log(\frac{\nu_{1}S_{\nu}(\lambda_{1})}
                         {\nu_{2}S_{\nu}(\lambda_{2})}) 
\end{equation}

\noindent
similar to that used by Sivagnanam et~al. (1990).
	
	The selected objects were intended to cover the far right-hand side 
of the region labelled ``Mira-like'' by  Sivagnanam et~al. (1990) 
and the areas labelled LSDD and $R_2$ by 
David et~al. (1993a) for the OH/IR sources. 
These areas are respectively delimited as follows~:

\begin{eqnarray}
         -0.50<[25-12]<-0.40  \nonumber \\
         -1.30<[60-25]<-0.80 
  \label{eq:region 1}
\end{eqnarray}
\begin{eqnarray}
         -0.40<[25-12]<+0.00      \nonumber \\
         -1.20<[60-25]<-0.80 
  \label{eq:region 2}
\end{eqnarray}
\noindent
and
\begin{eqnarray}
         -0.10<[25-12]<+0.40  \nonumber \\
         -0.80<[60-25]<-0.20 
  \label{eq:region 3}
\end{eqnarray}

These areas are covered by the regions labelled IIIa 
(i.e., {\em variable stars with more evolved O-rich circumstellar shells}), 
IIIb (i.e.,{\em variable stars with thick O-rich circumstellar shells}) and 
IV (i.e., {\em variable stars with very thick O-rich circumstellar shells}) by 
van~der~Veen \& Habing (1988) and van~der~Veen \& Rugers (1989). \\

The observations by David et~al. (1993a) were done at 1612 and 
1667~MHz, and those of  Sivagnanam et~al. (1990) at 1667~MHz only. 
Both sets of observations were carried out with the {Nan\c cay} radiotelescope 
with the same resolution as the present work, allowing a direct 
comparison of the profile shapes. Observations made by David et~al. 
(1993a) for the stars of interest were taken over a period of 5.5 years from 
mid-1986 to the early 1992. Those of Sivagnanam et~al. (1990) were all 
obtained in July 1986 except that of IRAS~20547$+$0247 which was obtained in 
May 1987.

\section{Results: General comments}  \label{section Results}

	Table~\ref{Table:evolution spectrale 1} presents the main 
characteristics of the sources of the study. Columns 1 to 10 give the name of 
the source, its galactic coordinates, its LRS class, its IRAS flux at 12, 25 
and 60~$\mu$m, its $[25-12]$ and $[60-25]$ colour indices and finally the 
name under which it is also known. The table was divided such that the three 
parts correspond to the three regions ``Mira-like'', LSDD and $R_2$ defined 
by the colour indices given in (\ref{eq:region 1}), (\ref{eq:region 2}) and 
(\ref{eq:region 3}) respectively.

%
\noindent
\begin{table*}
\caption{Galactic positions, IRAS flux at 12, 25 and 60~$\mu$m for the sources 
	 of the study} 

\label{Table:evolution spectrale 1}
                \scriptsize
\begin{tabular*}{13.4cm}{@{}@{}rrrrrrrrrl}
\hline
 Iras name & $l$   & $b$   & LRS & S$_{12\mu m}$ & S$_{25\mu m}$ & 
 S$_{60\mu m}$ & [25-12] & [60-25] & other name \\
            & deg.   & deg. & class &    Jy       &    Jy         &
      Jy       &           &         &     \\
\hline
15226$-$3603 & 334.7 &     17.0 & 28 &  166.40 &  121.10 &  16.58 & $-$0.457 & 
	    $-$1.244 &  \\
15262+0400 &     8.1 &     45.8 & 27 &   46.15 &   30.87 &   4.70 & $-$0.493 & 
	    $-$1.198 &  \\
20396$-$0826 &  38.3 &  $-$28.4 & 69 &   11.71 &    9.23 &   1.60 & $-$0.422 & 
 	    $-$1.141 &  XX~Aqr \\
20547+0247 &    51.4 &  $-$26.1 & 32 &   45.52 &   33.76 &  10.02 & $-$0.449 & 
	    $-$0.908 &  U~Equ \\
22525+6033 &   109.2 &      1.1 & 24 &  112.10 &   92.30 &  17.05 & $-$0.403 & 
	    $-$1.114 &  \\
\hline
17482$-$2824 &     1.1 & $-$0.8 & 42 &  428.70 &  480.40 &  109.10 & $-$0.269 & 
	  $-$1.024 &  OH 1.1$-$0.8 \\
17505$-$3143 &   358.5 & $-$2.9 & 34 &   47.60 &   69.51 &   23.82 & $-$0.154 & 
	  $-$0.845 &  \\
19075+0921 &  43.4 &     0.2 & 12 &  133.00 &  163.70 &   54.91 & $-$0.229 & 
	  $-$0.855 &  \\
19161+2343 &  57.1 &     5.1 & 31 &  112.20 &  137.20 &   31.04 & $-$0.231 & 
	  $-$1.026 &  \\
19190+1128 &  46.6 &  $-$1.3 & 32 &   27.13 &   36.60 &   10.22 & $-$0.189 & 
	  $-$0.934 &  \\
19244+1115 &  47.1 &  $-$2.5 & 28 & 1346.00 & 2314.00 &  717.70 & $-$0.083 & 
	  $-$0.889 &  IRC~10420 \\
\hline
17220$-$2448 &   1.0 &     6.1 &    &    5.35 &    9.06 &    3.84 & $-$0.090 & 
	 $-$0.753 &  \\
17317$-$3331 & 354.9 &  $-$0.5 & 79 &  104.00 &  291.50 &  234.50 &  0.129 & 
	 $-$0.475 &  \\
17385$-$3332 & 355.6 &  $-$1.7 &    &    2.88 &   13.25 &   10.01 &  0.344 & 
	 $-$0.502 &  \\
17392$-$3319 & 355.9 &  $-$1.7 & 37 &   19.12 &   37.29 &   16.49 & $-$0.029 & 
	 $-$0.735 &  \\
17411$-$3154 & 357.3 &  $-$1.3 &    & 1262.00 & 2723.00 & 1365.00 &  0.015 & 
	 $-$0.680 &  \\
17418$-$2713 &   1.4 &     1.0 & 79 &   14.95 &   51.28 &   45.94 &  0.217 & 
	 $-$0.428 &  OH~1.3+1.0 \\
17550$-$2120 &   8.0 &     1.4 &    &    5.38 &   21.08 &   30.81 &  0.274 & 
	 $-$0.215 &  OH~8.0+1.4 \\
18033$-$2111 &   9.1 &  $-$0.2 &    &    9.75 &   20.84 &    9.20 &  0.011 & 
	 $-$0.735 &  \\
18135$-$1456 &  15.7 &     0.8 & 79 &   31.02 &  124.40 &  157.60 &  0.284 & 
	 $-$0.277 &  OH~15.7+0.8 \\
18198$-$1249 &  18.3 &     0.4 &    &   15.56 &   42.54 &   33.91 &  0.118 & 
	 $-$0.479 &  OH~18.3+0.4 \\
18257$-$1000 &  21.5 &     0.5 & 39 &   46.32 &  120.50 &  115.90 &  0.096 & 
	 $-$0.397 &  OH~21.3+0.4 \\
18348$-$0526 &  26.5 &     0.6 &    &  359.80 &  633.80 &  463.00 & $-$0.073 & 
	 $-$0.517 &  OH~26.5+0.6 \\
18432$-$0149 &  30.7 &     0.4 &    &   25.06 &   52.32 &   48.09 &  0.001 & 
	 $-$0.417 &  OH~30.7+0.4 \\
18460$-$0254 &  30.1 &  $-$0.7 &    &  111.10 &  279.90 &  806.90 &  0.083 & 
	   0.080 &  OH~30.2$-$0.7 \\
18488$-$0107 &  31.9 &  $-$0.4 &    &   16.45 &   42.98 &   44.03 &  0.098 & 
	 $-$0.370 &  OH~32.0$-$0.5 \\
19065+0832 &  42.6 &     0.1 & 39 &   20.22 &   51.52 &   31.18 &  0.087 & 
	 $-$0.598 &  OH~42.6+0.0 \\
19254+1631 &  51.8 &  $-$0.2 & 39 &   16.57 &   44.67 &   36.27 &  0.112 & 
	 $-$0.471 &  OH~51.8$-$0.2 \\
19352+2030 &  56.4 &  $-$0.3 & 03 &   42.89 &   97.88 &   43.12 &  0.040 & 
	 $-$0.736 &  OH~56.4$-$0.3 \\
22177+5936 & 104.9 &     2.4 & 38 &  123.20 &  228.90 &   90.65 & $-$0.050 & 
	  $-$0.782 &  OH~104.9+2.4\\
\hline
\end{tabular*}
                        \normalsize
\end{table*}    


	The results concerning the spectral profile changes of the individual 
sources in the three regions of concern are presented in 
Appendix~\ref{appendix A}.  \\

Figure~\ref{fig:profil spectral evolution} presents the spectral profiles of 
each source at 1665 and 1667~MHz in both circular polarizations. 
As IRAS~18198$-$1249, IRAS~18257$-$1000, IRAS~18488$-$0107, IRAS~19065$+$0832, 
IRAS~22177$+$5936 and IRAS~19075$+$0921 had already been detected before the 
work of David et~al. (1993a), their spectra were not displayed by the 
previously cited authors. The spectral profiles of the 5 first cited sources
are given in Dickinson \& Turner (1991) and that of IRAS~19075$+$0921 
is given in Le~Squeren et~al. (1992). 
Therefore, for these 6 sources, the spectral shape changes were determined 
from the comparison of our observations with those of the two papers 
mentioned, while for the other sources, comparison has been made with those of 
David et~al. for the OH/IR stars and Sivagnanam et~al. (1990) for 
the Miras.
	All the sources of the $R_2$ region were also observed at 1612~MHz 
by  David et~al. (1993b) and quite a few by te~Lintel et~al. 
(1991). Three, out of the six sources of the LSDD region were observed at 
1612~MHz by te~Lintel et~al. (1991). Finally, all the sources of the 
``Mira-like'' region were observed at 1667 and 1612~MHz by 
Sivagnanam et~al. (1990). \\

	Tables~\ref{Table:evolution spectrale 65} and 
\ref{Table:evolution spectrale 67} present the results of the spectral 
measurements done at 1665 and 1667~MHz respectively. As for 
Table~\ref{Table:evolution spectrale 1}, these tables were divided into three 
parts corresponding to the three regions of concern. Columns 1 to 16 give the 
name of the source, a comment about its profile shape in the case of a non 
``standard'' double-peaked profile, the velocity of the blue- 
and red-shifted peaks and the velocity of any inter-peak component in both 
circular polarizations, the stellar and expansion velocities inferred, the 
flux intensity of the blue- and red-shifted peak and that of any  
inter-peak component in both circular polarizations 
(i.e., measured respectively at the velocity given in column 3 to 8) 
respectively. For the profile shape we distinguish between 4 
different categories~: 
``standard'' type spectrum but exhibiting inter-peak signal (``s''), plateau 
profile spectrum (``p''), spectrum with more than two well detached peaks 
(i.e., ``standard'' profile plus an internal feature, ``ci'' where 
``internal'' means in between the two standard peaks) and finally spectrum 
showing only a single peak centred at or about the stellar velocity (``c''). 
For the two last categories, the characteristics of the signal are given in 
columns~7, 8 and 15, 16. \\

	David et~al. (1993a) used the 1985 IRAS fluxes while we used the 
corrected 1987 fluxes. Thus, IRAS~18460-0254 which was initially in the $R_2$
region of David et~al. is now outside this box due to a notable 
difference in its $[60-25]$ colour index when calculated with or without 
correction (cf. Fig.~\ref{diagcoul}). Nevertheless, its $[25-12]$ colour 
index is hardly affected by the application of the IRAS correction. Since 
this index is the most relevant in terms of mass loss and shell thickness 
dependence, we decided to incorporate this source in the statistics of the 
$R_2$ region. 
 
\section{Discussion}  \label{section Discus.}

\subsection{The ``Mira-like'' sample}

The area corresponding to the OH-emitter Miras contains five objects which,
according to their location in the colour-colour diagram, have the
thickest shells of their category. 
None of the five sources exhibits a standard spectral profile
either at 1665 or 1667~MHz and only one source shows a double-peaked
profile characterized by two unevenly broad and spiky peaks. This source
has the smallest [25$-$12] colour index value of the sample. All the 
sources show quite irregular spectral profile shapes. 
This tendency is also clearly observable in the spectral profiles obtained 
by Sivagnanam et~al. (1990, their Fig.~2). Such a strong departure from 
the double-peaked spectral profile indicates a strong heterogeneity in the 
physical and dynamical conditions ruling over the whole OH shell even 
though those Miras have quite thick shells.

All the sources show OH variability. Among the five sources selected, four of 
them were primarily successfully detected in the satellite line at 1612~MHz 
and we detected 1665~MHz emission from all of them. For most of the sources of 
the ``Mira-like'' sample positively detected by Sivagnanam et~al. (1990)
as OH-emitters, when 1612~MHz emission was detected its intensity was 
comparable to or greater than that at 1667~MHz. Such a trend in the 
strengthening of the 1612~MHz emission over that of the main lines is expected 
when the shell is getting thick enough (cf. Lewis 1989; David {\it et~al.} 
1993b; Habing 1996). The sample selected here 
(i.e., $-0.50<$[25$-$12]$<-0.40$) is such that it fulfils 
the physical conditions for the 1612~MHz and the main-line emissions to be of 
similar strength. IRAS~15226$-$3603 was one of the rare sources not detected 
positively at 1667~MHz, nor did we detect any 1665~MHz emission from it. 
Since this source was detected positively by Sivagnanam {\it et~al.} earlier, 
our non-detection is most likely due to the fact that the source was at 
its OH minimum at the time of our observations. Finally, one source 
(IRAS~22525$+$6033) appeared to be strongly polarized in both main lines. This 
source exhibits the strongest 1665~MHz emission and the greatest expansion 
velocity of its group: about 10~km~s$^{-1}$. Thus, this source is located in 
the narrow zone where both Miras and OH/IR stars can be found in the 
``period$-$OH expansion velocity'' diagram  as given by 
Sivagnanam {\it et~al.} (1989). 
\subsection{The OH/IR sample of the LSDD region}

This sample of six sources presents the highest rate of double-peaked spectral 
profiles at 1667~MHz ($>$50\%) with very faint or non-existent inter-peak 
emission. Four of the six sources (i.e., 67\%) were detected at 
1665~MHz at the time of the observations. Nevertheless, none of the sources 
shows a ``classical'' double-peaked profile at that frequency. In addition, 
the 1665~MHz emission for the sources of this sample is very faint. Only 
IRAS~19075$+$0921 shows a significant flux density at 1665~MHz 
(i.e., comparable to that observed at 1667~MHz). 
This particular source also shows strong polarization in both main lines and
has the smallest expansion velocity of the whole sample, with a value of 
7.7~km~s$^{-1}$. Consequently, like IRAS~22525+6033, IRAS~19075$+$0921 
is located in the zone where Miras and OH/IR stars coexist in the 
``period$-$OH expansion velocity'' diagram of 
Sivagnanam {\it et~al.} (1989). \\

\subsection{The OH/IR sample of the $R_2$ region}

The $R_2$ sample is composed of nineteen OH/IR stars. 84\% of the stars of 
this group show at least two peaks in their 1667~MHz spectral profile with 
only 31.5\% showing a standard double-peaked profile (i.e., with no inter-peak 
signal). The percentage of stars having at least two peaks at 1665~MHz is only 
48\%. Five sources (i.e., 26\%) exhibit a single-peaked spectral profile and 
five other sources did not show any 1665~MHz emission at all. 26\% of the 
sources that were positively detected at 1665~MHz show an internal group of 
components with an intensity comparable or superior to that of the red- or 
blue-shifted peaks. 
 
Among the whole set of the double-peaked profiles of this group of thicker 
OH/IR stars, nine sources (i.e., 48\%) show an inter-peak signal at 1667~MHz. 
Sun (1991) has modelled the evolution of the 1612~MHz double-peaked 
profile when the mass-loss rate varies with time. The author shows that the 
inter-peak signal at 1612~MHz increases with the mass-loss rate. Our results 
show that this property also applies to the 1667~MHz emission. Also, the 
existence of an inter-peak signal is related to the degree of saturation 
(Fix 1978). 
Therefore, the increased number of inter-peak signals observed in the 
profiles of these thicker OH/IR stars is quite probably a signature of 
greater maser saturation at 1667~MHz.

\subsection{Spectral profile shape as a function 
		of the shell thickness} \label{spectral profile and thickness}

Considering the three groups together, it stands out that the spectral 
profiles with inter-peak signal as strong as that of the red- or 
blue-shifted peaks are only observed for OH/IR sources at 1665~MHz. 
OH/IR stars with inter-peak signal at 1667~MHz are found in both the LSDD and 
the $R_2$ regions. Except IRAS~19352$+$2030 which shows very faint blue- and 
red-shifted peaks and a very strong inter-peak component in both main lines, 
the sources show a faint 1667~MHz inter-peak signal which can be 
slightly polarized. Spectral profiles showing a well-detached inter-peak 
component at 1665~MHz arise from OH/IR stars belonging to the upper 
part of the $R_2$ region, that is, from the stars which have a very thick 
shell. The ``plateau'' profiles are only observed at 1665~MHz, but in all 
three groups. It can be particularly pronounced for some objects 
(cf. IRAS~17411$-$3154).
Such a profile shape generally attests to significant acceleration (i.e., 
Chapman \& Cohen, 1985 for a velocity gradient $\epsilon \simeq 1$). 
This confirms that the 1665~MHz emissive zone indeed originates deeper 
inside the OH shell than the 1667~MHz zone. \\

\begin{figure}
   \begin{center}
        \epsfig{file=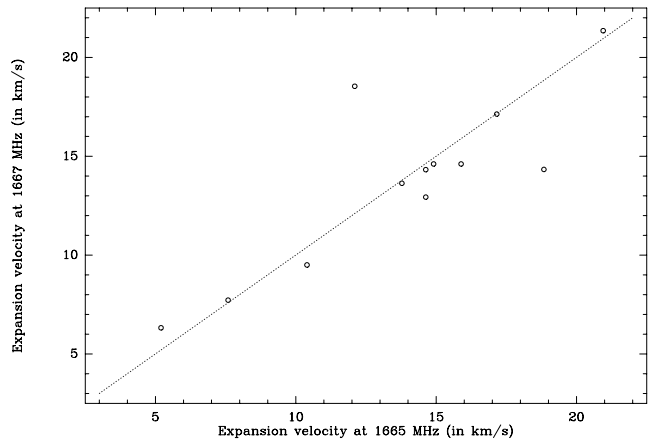,width=6.5cm,angle=-90}
   \end{center}
   \caption{Expansion velocity at 1667~MHz versus that at 1665~MHz}
   \label{fig:vitcarevol6567}
\end{figure}

Figure~\ref{fig:vitcarevol6567} shows the correlation between the expansion 
velocity inferred at 1665~MHz and that at 1667~MHz. This diagram uses 
the measurements from all the sources that showed at least two peaks in both 
main lines except IRAS~19244$+$1115 for which the expansion velocities 
inferred have to be taken with caution (cf. Appendix~\ref{sub OHIR region2}). 
The total number of such sources is 12, that is about 41\% of 
the whole sample of 29 objects for this statistic. This corresponds 
respectively to 40, 40 and 42\% of the ``Mira-like'', LSDD and $R_2$ samples. 
The dotted line in the figure is the locus for  
$V_{\rm 1665\,MHz}$=$V_{\rm 1667\,MHz}$. Most of the sources have a similar 
expansion velocity in both main lines. At the location of the OH shell, some 
10$^{16}$~cm away from the central star, turbulence of 1-2~km~s$^{-1}$ is 
expected (Deguchi, 1982). 
This could result in the broadening of the line wings and therefore explain 
most of the cases where 1665~MHz profiles overshoot those of the 1667~MHz line 
by such a value.
One source, IRAS~17392$-$3319 from the $R_2$ region, shows an expansion 
velocity at 1665~MHz greater by 4~km~s$^{-1}$ than that inferred at 1667~MHz. 
This may be due to the faintness of the signal at 1665~MHz 
leading to a bigger uncertainty in the exact location of the peak maxima.  \\
 
Only four sources show an intensity at 1665~MHz greater than that at 
1667~MHz. These sources belong to the ``Mira-like'' and LSDD regions. None of 
the sources of the $R_2$ region show this peculiarity. We have investigated 
whether there was any correlation between the intensity observed in the 
main lines and the thickness of the shell. Taking all the sources showing both 
1665 and 1667~MHz emission, we have found that there is a relation between 
the ratio of the intensity observed in the two main lines and the [25$-$12] 
colour index. 
This relation is shown Fig.~\ref{fig:evolimax6567}, where the dotted line 
shows the linear regression over all the sources except the three for 
which Intensity$_{1667} >>$ Intensity$_{1665}$ 
(plotted with filled symbols). The linear regression is given by~: 

\begin{equation}
 	\label{regression lineaire}
 {   Int(1667/1665)_{i}=6.54 \times [25-12]_i +4.40
 }
\end{equation}
 
\noindent
where $Int(1667/1665)_{i}$ is the ratio of the maximum intensity measured at 
1667 and 1665~MHz for the source ``i''  
and [25$-$12]$_i$ is its colour index. 

Three sources do not lie on the relation given by 
Eq.~({\ref{regression lineaire}). 
We do not think that this could be due to variability. 
Indeed, for one of the sources (IRAS~18432$-$0149) the 1665~MHz emission 
faintness ($\sim 2 \sigma$) could be responsible for the high $Int(1667/1665)$ 
found but such an explanation is not plausible for the two other sources which 
showed reasonably strong 1665~MHz emission at the time of the observations. 
On the other hand, it is noteworthy that the three sources not following the 
general trend have similar [25$-$12] indices~: [0;0.1]. 

Equation~(\ref{regression lineaire}) shows that when the shell 
becomes thicker, the 1667~MHz emission is favoured in comparison with that at 
1665~MHz. Since the expansion velocity is about the same at 1665 and 1667~MHz 
it is expected that the maser emission from both main lines should be found 
at similar distances from the star and possibly share the same regions 
of the OH shell. This hypothesis is supported by the maps of late-type stars
obtained in VLBI (cf. Chapman et~al. 1991, 1994; Szymczak et~al. 
1999). 
Therefore, a plausible explanation for such a trend is an increase of  
competitive gain in favour of the 1667~MHz line when the shell becomes 
thicker. 

\begin{figure}
   \begin{center}
        \epsfig{file=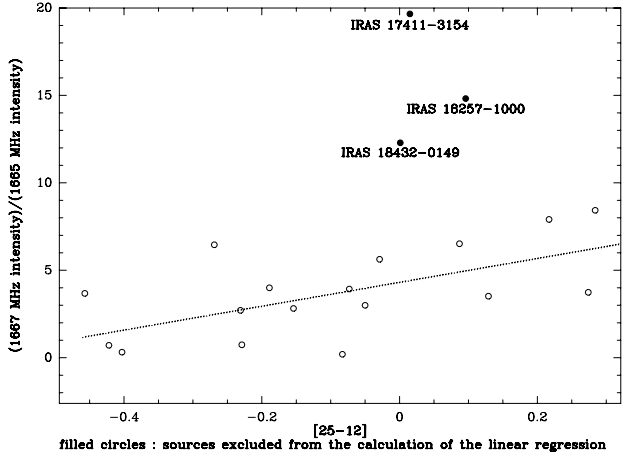,width=6.5cm,angle=-90}
   \end{center}
   \caption{(1667 MHz)/(1665 MHz) intensity ratio versus the colour index 
	[25-12]}
   \label{fig:evolimax6567}
\end{figure}

	This increase of the ratio of the 1667~MHz emission over that at 
1665~MHz is in agreement with the fact that 1667~MHZ emission becomes 
relatively stronger and eventually dominant in the post-AGB stage 
(Zijlstra et~al. 1989, 2001). The dominance of the 1667~MHz emission at 
this stage of the stellar evolution is supported by theoretical studies 
(Field 1985, Field \& Gray 1988).

\subsection{The Mira-OH/IR sequence}

The Miras-OH/IR sequence seen in the IRAS colour-colour diagram results from 
the combination of various relations existing amongst the physical parameters 
of the central star and the circumstellar envelopes~: (1) between the 
mass-loss rate and the period (Vassiliadis \& Wood 1993), (2) between the 
period and the expansion velocity (Dickinson \& Chaison 1973, Le~Squeren 
et~al. 1979), (3) between the OH luminosity and the flux at 25~$\mu$m 
(Sivagnanam et~al. 1989), (4) between the OH radius and the OH 
luminosity (Baud \& Habing 1983), (5) between the size of the maser regions 
and the mass-loss rate (predicted for the H$_2$O masers by 
Cooke \& Elitzur 1985 and observed by Cohen 1987, Lane et~al. 1987 and 
Yates \& Cohen 1994) and (6) between the mass-loss rate and the optical 
depth at 10~$\mu$m (Bedijn 1988). 
This can by summarized as follows~: the increase of the pulsation period, the 
expansion velocity, the radius of the circumstellar shell and the luminosity 
are all function of the mass-loss rate and the opacity of the shell.

The sequence observed in the colour-colour diagram may be either 
evolutionary or related to the stellar mass. If it is an evolutionary 
sequence the star should experience an increase of its mass-loss rate 
with time. If it is a mass sequence the mass loss should be connected 
somehow to the initial mass. Volk \& Kwok (1988) studied the IR spectral 
evolution of AGB stars. Their study shows that AGB stars that have a 
silicate signature at 10~$\mu$m in absorption are mainly stars with an initial 
mass on the main sequence greater than 3~M$_\odot$. This implies that the 
shell thickness depends on the initial mass, since only the stars massive 
enough will have a thick shell i.e. will become OH/IR stars. Also, through a 
study of the galactic distribution of OH/IR stars according to their expansion 
velocity, Likkel (1989) found that the reddest OH/IR objects tend to have the 
highest initial masses. 
This latter observational result is in agreement with earlier work 
by Baud (1981) who showed the existence of a relation between the expansion 
velocity and the initial mass of the central star. \\

We found that IRAS~22525$+$6033 and IRAS~19075$-$0921 belonging to the 
``Mira-like'' group and to the LSDD group respectively show very similar 
characteristics~: 1665~MHz flux comparable to that at 1667~MHz, strong 
polarization over the whole velocity range of the spectrum in both main lines, 
similar OH expansion velocity values and [25$-$12] indices. This puts these 
two objects in the location where Miras and OH/IR stars coexist both in the 
``period-OH expansion velocity'' diagram and the colour-colour diagram, but 
they have a completely unusual set of spectral characteristics in comparison 
with the rest of the stars with similar expansion velocities and IRAS [12-25] 
fluxes.
In particular, the very strong polarization observed over the whole spectral 
profile is clearly a sign of perturbation undergone throughout the whole OH 
circumstellar shell, which we interpret to be a signature that these objects 
are in the transition phase Mira~$\rightarrow$~OH/IR star. Indeed, similar 
polarization characteristics in the OH shell have been observed 
(Bains et~al. 2003, Etoka \& Diamond 2004) attesting to the 
importance of the role played by the magnetic field (in particular in the 
envelope shaping) in the transitional stages of evolved stars.
Bearing in mind the existing relations between the initial mass and some 
of the actual physical properties of AGB stars such as their shell thickness 
or their expansion velocity, it is very likely that the IRAS colour-colour 
sequence is related to both the initial mass of the object and its 
evolutionary stage. A likely hypothesis is that the object arrives more or 
less high on the sequence and evolves through a part of it before expelling 
its shell entirely via an increasing mass loss.

\section{Conclusion}  \label{section Conclu.}

Acceleration and turbulence can strongly alter the spectral profile shapes 
of circumstellar OH masers. 
This is well illustrated by the variety of spectral profiles observed for the 
``Mira-like'' sample. When the shell gets thicker and less turbulent, the 
standard double-peaked profile settles. At that stage, saturation is the main 
cause of spectral profile alteration at 1667~MHz, causing an inter-peak signal 
to appear while plateau-shape or strong inter-peak components observed in 
the spectral profiles at 1665~MHz still attest to the presence of 
acceleration. Even though the spectral profile characteristics show that the 
1665~MHz emission can be found deeper in the OH shell than the 1667~MHz 
emission, the expansion velocity is found to be about the same in both 
main lines, taking into account a possible velocity turbulence of 
1-2~km~s$^{-1}$ at the location of the main-line maser emission. It is then 
expected that the maser emission from both main lines should be found at 
similar distances from the star and possibly share the same regions of the OH 
shell. Also, the intensity ratio 1667/1665 has been found to increase with the 
shell thickness. This can be explained as an increase of competitive gain 
in favour of the 1667~MHz when the shell is becoming thicker. Finally, a Mira 
and an OH/IR star from the lower part of the colour$-$colour diagram show 
quite strong polarization while the rest of the sources are weakly 
polarized. Strong polarization can be the sign of dramatic changes in the 
shell itself. We suspect that the strong polarization observed in those two 
particular objects may be linked with their being in the phase of transition
Mira~$\rightarrow$~OH/IR star. 


\newpage
\appendix
\section{Remarks on individual sources} \label{appendix A}

\subsection{``Mira-like''} \label{sub Mira}

{\bf IRAS~15226$-$3603~:} We did not detect any emission from this source 
either at 1665 or at 1667~MHz in March 1994 although Sivagnanam et~al. 
(1990) clearly detected a signal at 1667~MHz in July 1986 with an intensity 
of 0.46~Jy for the strongest spectral profile component. At the time of 
the observations by Sivagnanam et~al. this source was not detected at 
1612~MHz, thus this Mira is certainly a type~I. \\

\noindent
{\bf IRAS~15262$+$0400~:} At the time of our observations this source showed 
an intensity at 1667~MHz seven times greater than that observed by Sivagnanam 
et~al. (1990) in July 1986. Nevertheless, the intensity ratio 
$I_{\rm red \; peak}$/ $I_{\rm blue \; peak}$ (where ``red peak'' and 
``blue peak''stand for the red-shifted and blue-shifted emission respectively) 
stayed the same. We detected faint non-polarized emission ($\simeq$~0.2~K)
at 1665~MHz internal in velocity to that observed at 1667~MHz (where 
``internal (in velocity)'' should be understood as 
defined in Sect.~\ref{section Results}, meaning in this case that the signal 
is located between the blue and red peaks observed at 1667~MHz). \\

\noindent
{\bf IRAS~20396$-$0826~:} This identified Mira, also known as XX~Aqr, has 
an optical period of 334~days (Kholopov et~al. 1985, {\it GCVS}). It was 
classified by  Sivagnanam et~al. (1990) as a ``peculiar object'' due to the 
difference between the spectral profile shape observed at 1612 and that 
observed at 1667~MHz. The 1612~MHz emission shows a double-peaked profile 
covering a velocity range of 6~km~s$^{-1}$. The 1667~MHz emission observed by 
Sivagnanam et~al. is internal in velocity to that observed at 1612~MHz 
and has a plateau-shaped profile. Our observations show that the 1665~MHz 
emission is internal in velocity to that observed at 1667~MHz with a single 
peak centred on the stellar velocity. A strong acceleration ($\epsilon >1$)
at the location of the 1665~MHz emission could produce such a spectral profile
shape (cf. Chapman \& Cohen 1985), strengthening the
hypothesis of a velocity gradient in the OH shell of this source as 
proposed by Sivagnanam et~al.. \\

\noindent
{\bf IRAS~20547$+$0247~:} This source, also known as U~Equ, was indexed 
by Sivagnanam et~al. as the Mira having the smallest expansion velocity 
with a value of 2.4~km~s$^{-1}$ (cf. their Table~1). The 1667~MHz intensity of 
this source has substantially dropped. The blue-shifted peak which was already 
very faint at the time of the observations by Sivagnanam et~al. 
in May 1987 is 
no longer detected in our observations. We detected a very faint 1665~MHz 
signal only in the red-shifted peak centred at the same velocity as the 
emission we observed at 1667~MHz. \\

\noindent
{\bf IRAS~22525$+$6033~:} Sivagnanam et~al. classified this source as a 
``peculiar object'' due to its spectral profile at 1667~MHz. Our observations 
indeed confirm the non-standard profile shape in both main lines, with 
multiple components quite well detached most of which are strongly right-hand 
polarized (cf. Fig~\ref{fig:IRAS22525+6033 D-G} showing the 
Stokes $V$ spectra for both main lines). The resulting spectral profiles in 
the two circular polarizations are therefore very different 
(cf. Fig~\ref{fig:profil spectral evolution}). Sivagnanam et~al. 
inferred an expansion velocity of 4.4~km~s$^{-1}$. From our observations we 
infer an expansion velocity more than twice that value~: 9$-$10~km~s$^{-1}$. \\

\begin{figure}
   \begin{center}
        \epsfig{file=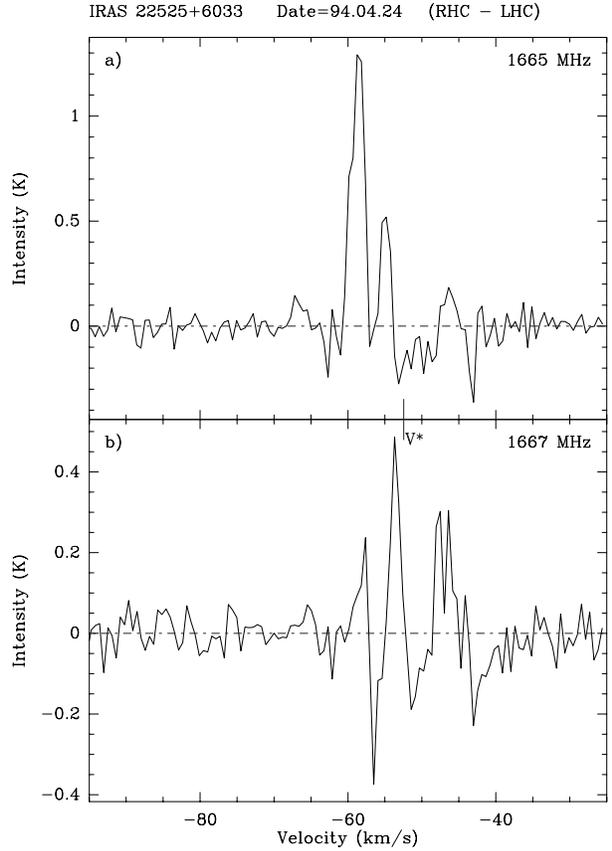,width=8.5cm}
   \end{center}
   \caption{Stokes $V$ spectra of IRAS~22525+6033: {\bf a)} at 1665~MHz, 
		{\bf b)} at 1667~MHz. V* is the stellar velocity inferred
		from the present work}
   \label{fig:IRAS22525+6033 D-G}
\end{figure}

\subsection{OH/IR stars from the LDSS region} \label{sub OHIR region2}

{\bf IRAS~17482$-$2824~:} Emission at 1667 and 1665~MHz is present. The 
velocity location for the red- and blue-shifted peaks is about the same in 
both lines but the 1667-MHz red-shifted peak emission overshoots that at 
1665~MHz by about 1~km~s$^{-1}$. Even though there is a separation of two 
years between the 1667-MHz observations made by David et~al. (1993a) 
and ours, there is no significant change in the spectral profile and 
intensity ratio $I_{\rm red \; peak}$/ $I_{\rm blue \;peak}$. \\

\noindent
{\bf IRAS~17505$-$3143~:} The comparison between the 1667-MHz observations 
made by David et~al. (1993a) and ours showed a noticeable change in 
both the flux and the intensity ratio 
$I_{\rm red \; peak}$/ $I_{\rm blue \; peak}$ 
between the two epochs of observations. We did not detect any 1665~MHz 
emission. This source was observed repeatedly at 1612~MHz. In particular, 
te~Lintel et~al. (1991) observed it in November 1985. They found an 
expansion velocity of 15~km~s$^{-1}$ with the blue- and red-shifted peaks 
located at $-12.2$ and $+19.2$~km~s$^{-1}$ respectively (cf. their Table~2). 
From our 1667~MHz observation we found the blue- and red-shifted peaks to be 
centred at $-12.9$ and $+20.3$~km~s$^{-1}$ respectively and an expansion 
velocity of 16.6~km~s$^{-1}$ in agreement with David et~al. (1993a). 
This implies that the 1667~MHz red-shifted peak overshoots that at 1612~MHz 
by about 1~km~s$^{-1}$. \\

\noindent
{\bf IRAS~19075$+$0921~:} The spectral profile of this source shows 
detached multi-components in both main lines leading to a 
non-standard shape, particularly noticeable at 1665~MHz. 
Like IRAS~22525$+$6033, this source shows a strong degree of polarization. 
Nevertheless, in this object the left-handed polarization 
prevails as can be seen in Fig.~\ref{fig:IRAS19075+0921 G-D} which displays 
the Stokes $V$ spectra in both main lines. As a result of the strong degree 
of circular polarization of some of the spectral components, the profile is 
quite different in both circular polarizations for both main lines. The 
1667~MHz spectrum profile obtained by Le~Squeren et~al. (1992) in 
January 1986 is similar to ours. The intensity ratio 
$I_{\rm red \; peak}$/ $I_{\rm blue \; peak}$ 
did not show any significant change bewteen the two sets of observations. 
We also found the same characteristic velocities as those given by Le~Squeren 
et~al. (1992, their Table~ 1). \\

\begin{figure}
   \begin{center}
        \epsfig{file=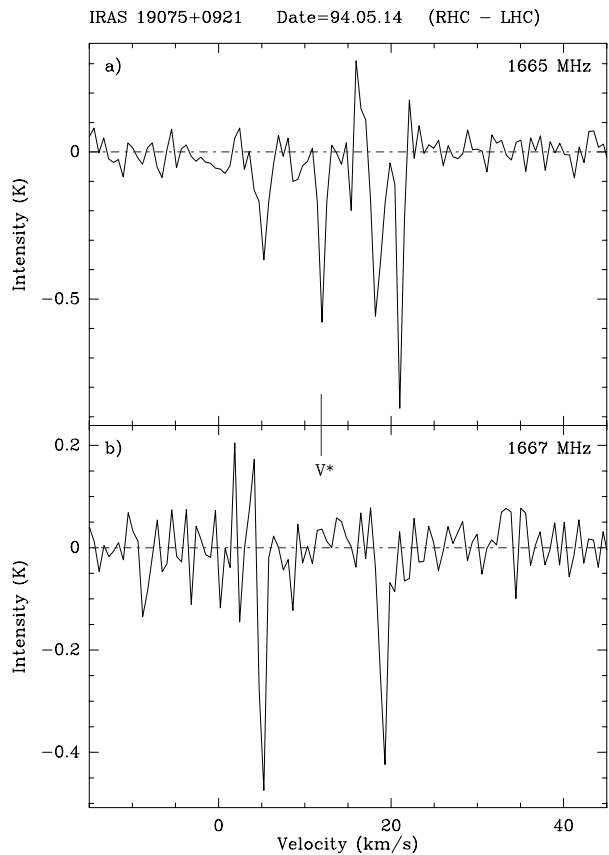,width=8.5cm}
   \end{center}
   \caption{Stokes $V$ spectra of IRAS~19075+0921: {\bf a)} at 1665~MHz, 
		{\bf b)} at  1667~MHz}
   \label{fig:IRAS19075+0921 G-D}
\end{figure}

\noindent
{\bf IRAS~19161$+$2343~:} We detected only a very faint and narrow emission 
in the red-shifted peak at 1665~MHz. The 1667~MHz red-shifted peak overshoots 
that at 1665~MHz by $\sim$1~km~s$^{-1}$. 
No significant change in the profile shape can be seen between the 
observations made by David et~al. (1993a) in May 1990 and ours. \\

\noindent
{\bf IRAS~19190$+$1128~:} This source shows very faint and asymmetric 
emission at 1667~MHz. In particular, the blue-shifted peak hardly reached 
a 2$\sigma$ detection. We do not notice any significant change in the 
profile shape between the observations made by David et~al. (1993a) 
in August 1986 and ours, performed 8 years later. We did not detect any 
emission at 1665~MHz. \\

\noindent
{\bf IRAS~19244$+$1115~:} This source, also known as IRC~10420, has quite a 
broad spectral profile. Because of the protocol of our observations
the red-shifted part of the spectra has been cut.
Indeed, if we refer to the observations made by David et~al. (1993a) at 
1667~MHz, we can clearly see the existence of a strong red-shifted component 
centred at +110~km~s$^{-1}$, outside our bandwidth coverage. Although it was 
covered by our bandwidth, we did not detect the most blue-shifted component 
centred at +41~km~s$^{-1}$ seen in the  profile published by David et~al.. 
That is most probably due to the ``band edge effects'' concerning the first 
and last 5 to 10 channels of the band.
Due to the inadequate protocol of observation for this particular source, the 
stellar and expansion velocities given in 
Tables~\ref{Table:evolution spectrale 65} and 
\ref{Table:evolution spectrale 67} should be treated with caution. This source
was not used in the statistical study on the expansion velocity presented in
Sect.~\ref{spectral profile and thickness}.

\subsection{OH/IR stars from the  $R_2$ region} \label{sub OHIR region3}

{\bf IRAS~17220$-$2448~:} We did not detect any emission at 1667~MHz, but a 
faint, narrow left-hand polarized component centred at V=$+45.3$~km~s$^{-1}$ 
was detected at 1665~MHz. Considering the velocity characteristics 
given by  David et~al. (1993a, V$_*$=+39.4~km~s$^{-1}$ and 
V$_{exp}$=15.4~km~s$^{-1}$) this component should be an OH maser signature of 
an inner region belonging to the back part of the shell. \\

\noindent
{\bf IRAS~17317$-$3331~:} The 1667~MHz emission of this source exhibits 
an inter-peak signal. At 1665~MHz we observed an internal group of components 
between the two asymmetrical red- and blue-shifted peaks. This internal peak 
is centred about the stellar velocity and its intensity is of the same order 
as that observed for the blue-shifted peak. This source was also observed at 
1612~MHz by te~Lintel et~al. (1991) and by David et~al. (1993a). 
From our observations in the main lines, we found the same stellar velocity 
as David et~al. which is in agreement with that given by te~Lintel 
et~al. considering their spectral resolution. 
At 1612~MHz the spectral profile does not have any noticeable particularity. 
Both te~Lintel et~al. and David et~al. classified it in 
the ``standard double-peaked profile'' category. \\

\noindent
{\bf IRAS~17385$-$3332~:} Both main lines exhibit faint emission. 
The blue-shifted peak is not detected at 1667~MHz and only just 
detected at 1665~MHz. At 1665~MHz, we clearly observe an internal peak 
centred on the stellar velocity, and the bluest narrow component we detected 
is centred at the same velocity as the 1612~MHz blue-shifted peak as given 
by te~Lintel et~al. (1991) and  David et~al. (1993b). \\

\noindent
{\bf IRAS~17392$-$3319~:} We did not detect any emission at 1665~MHz. The 
1667~MHz emission shows a standard double-peaked profile. The profile and 
intensity have not shown any noticeable change between our observations 
carried out in 1994 and those performed by David et~al. 8 years 
earlier. \\

\noindent
{\bf IRAS~17411$-$3154~:} The 1665~MHz emission exhibits a non-standard 
profile shape where at least 8 well-defined components are observed, of which 
4 are coming from strong internal emission. Interestingly, the back part 
of the shell (i.e., V$>$V$_*$) shows quite strong right-handed polarization 
while the front part (i.e., V$<$V$_*$) is left-hand polarized
(Fig.~\ref{fig:IRAS17411$-$3154 D-G}). 
A similar phenomenon has already been observed in other sources (i.e., 
Ukita \& Le~Squeren 1984; Szymczak et~al. 2001), and
can be explained by the Cook mechanism when the magnetic field is aligned 
on the line of sight (Cook 1966, 1977). 
The 1667~MHz spectral profile is 
more typical although showing inter-peak emission. This source also exhibits 
strong 1612~MHz emission with intensity reaching a value as high as 
200-300~Jy 
(cf. Table~2 of te~Lintel et~al. 1991 and Table~1 of David et~al. 
1993b). The expansion velocity inferred at 1612~MHz is of the same order
as that inferred at 1667~MHz (i.e., 18.5~km~s$^{-1}$, 
Table~\ref{Table:evolution spectrale 67}). On the other hand, the most 
red-shifted part of the 1665~MHz emission was not detected, leading to a much 
smaller expansion velocity in that line. \\

\begin{figure}
   \begin{center}
        \epsfig{file=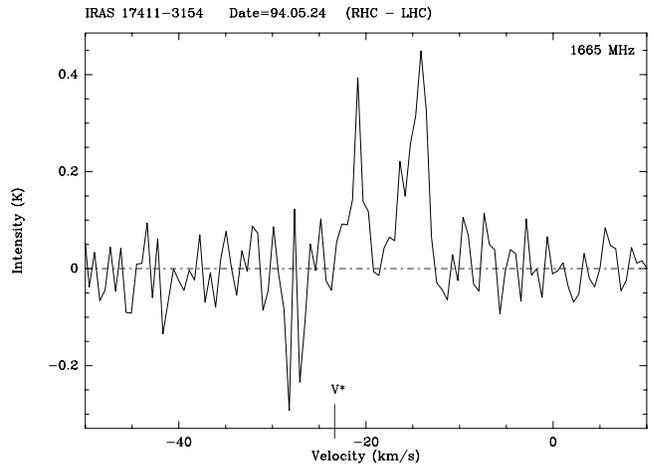,width=6.5cm,angle=-90}
   \end{center}
   \caption{Stokes $V$ spectrum of IRAS~17411$-$3154 at 1665~MHz.}
   \label{fig:IRAS17411$-$3154 D-G}
\end{figure}

\noindent
{\bf IRAS~17418$-$2713~:} This source shows a single component at 1665~MHz 
centred about V=$-21$~km~s$^{-1}$, that is about the mid-point between the 
stellar velocity and the 1667~MHz blue peak velocity. The expansion velocity 
inferred at 1667~MHz is more than 1~km~s$^{-1}$ greater than that inferred 
by David et~al. (1993b) and te~Lintel et~al. (1991) from the 
1612~MHz emission. This is due to an overshoot of the 1667~MHz blue peak 
over that at 1612~MHz. \\

\noindent
{\bf IRAS~17550$-$2120~:} Emission is observed in both main lines but with 
an inverse intensity ratio $I_{\rm red \; peak}$/ $I_{\rm blue \; peak}$. 
The 1667~MHz profile underwent a strong change of shape since 1987 
when a triple-peaked profile shape was observed  
(cf. Fig.~2 of David et~al. 1993a) while we detected only the 
two standard peaks. The extra internal component centred about 
V=$-17$~km~s$^{-1}$ detected by David et~al. vanished between 
the two epochs of observation. The nature of this event is unclear since 
the source has not been monitored. \\ 

\noindent
{\bf IRAS~18033$-$2111~:} It seems that the OH emission reported by
David et~al. (1993a, their Fig.~2 and Table~1) corresponds  
to two distinct sources~: one emitting in the velocity range 
[-130;-100]~km~s$^{-1}$ (the one we observed) and the second emitting in 
the velocity range [-70;-30]~km~s$^{-1}$. The 1612~MHz observations reported 
by te~Lintel et~al. (1991) and David et~al. (1993b) deal with the 
velocity range [-70;-30]~km~s$^{-1}$. 
To our knowledge no observation has 
been done at 1612~MHz in the velocity range [-130;-100]~km~s$^{-1}$. 
At the date of our observation neither 1667~MHz nor 1665~MHz emissions 
were detected while David et~al. (1993a) clearly reported 
emission at 1667~MHz. Thus, the source corresponding to the velocity 
range [-130;-100]~km~s$^{-1}$ is clearly variable. \\

\noindent
{\bf IRAS~18135$-$1456~:} The 1667~MHz spectrum of this source shows 
an asymmetric double-peaked profile. The 1665~MHz spectrum is internal 
to that at 1667~MHz and emission is observed over the whole velocity range
(leading to its classification as a plateau-shaped profile).
Actually, the two external components (i.e., the bluest and reddest 
Doppler shifted) are those with the faintest emission. In particular, if one 
excepts the strong component centred at V=$-9.4$~km~s$^{-1}$, the general 
shape of the profile resembles a gaussian. This could be the signature of an 
acceleration zone (cf. Chapman \& Cohen 1985). \\

\noindent
{\bf IRAS~18198$-$1249, IRAS~18257$-$1000, \\ IRAS~18488$-$0107 and 
IRAS~19065$+$0832~:} These four sources were observed by 
Dickinson \& Turner (1991) in both main lines. All four sources show a 
standard double-peaked profile at 1667~MHz. We did not notice 
any significant change in their spectral profiles. 
We did not detect any 1665~MHz either for IRAS~18198$-$1249 or for 
IRAS~18488$-$0107. We positively detected IRAS~18257$-$1000 and 
IRAS~19065$+$0832 at 1665~MHz. Dickinson \& Turner classified 
IRAS~18198$-$1249, IRAS~18257$-$1000 and IRAS~18488$-$0107 as sources for 
which the 1667~MHz emission overshoots that at 1612~MHz. \\

\noindent
{\bf IRAS~18348$-$0526~:} This source shows a strong inter-peak emission 
at 1667~MHz. At 1665~MHz we also noted a non-zero inter-peak emission whose  
shape resembles a very flat-topped gaussian centred on the stellar velocity. 
This kind of shape is similar to that presented by Bowers (1991) for an 
isotropic outflow in a spherical circumstellar shell with significant rotation.
This hypothesis would imply a non-negligible rotation component in the
external regions of the circumstellar shell considering the radius expected
for the OH formation (Goldreich \& Scoville 1976, Huggins \& Glassgold 1982).
This is very unlikely. Another more credible hypothesis regards the emission 
as the sum of two components. The first component, coming from an external OH 
region expanding radially with an acceleration such as $1>\epsilon>0$, would 
give the standard red- and blue-shifted peaks with possibly a faint inter-peak 
signal. The second component would come from an inner OH strongly accelerating 
zone with $\epsilon >1$, given the internal gaussian-like emission observed
(cf. Chapman \& Cohen 1985, their Fig.~3). This scenario is sustained by the 
1667~MHz profile shape which also shows an inter-peak signal, strengthening 
the hypothesis that some acceleration is still present at the location of the 
main-line maser emission. \\

\noindent
{\bf IRAS~18432$-$0149, IRAS~18460$-$0254 and \\ IRAS~19254$+$1631~:} The 
first two sources show faint emission at 1665~MHz barely above the noise 
threshold. The 1667~MHz spectra for all three sources did not show any 
significant change between the observations carried out 
by David et~al. (1993a) and ours. \\

\noindent
{\bf IRAS~19352$+$2030~:} At 1667~MHz we detected emission from three peaks 
centred at V=$-8.70$, $-1.4$ and $+7.0$~km~s$^{-1}$. The two most red- and 
blue-shifted peaks are very faint. Only the central peak at 
V=$-1.40$~km~s$^{-1}$ was detected by David et~al. (1993a). We also 
detected emission at 1665~MHz centred at V=$-1.40$~km~s$^{-1}$. This peak is 
significantly left-hand polarized (i.e., 15-20\%) in both main lines. This 
source was also observed at 1612~MHz by David et~al. (1993b). They only 
detected a single peak centred at V=$-3.22$~km~s$^{-1}$. \\

\noindent
{\bf IRAS~22177$+$5936~:} At 1667~MHz this source shows a
 double-peaked profile with strong emission between the the red- and 
blue-shifted peaks. At 1665~MHz only the blue-shifted peak is clearly 
detected. The red-shifted peak is barely above the noise threshold in  
right-handed polarization. The spectral profiles obtained by 
Dickinson \& Turner (1991) are similar to ours for both main lines. 
This source was also observed at 1612~MHz by te~Lintel et~al. (1991). 
The characteristic velocities they found are in agreement with ours.    

\section{Spectral line atlas } \label{appendix B}
\begin{figure*}
   \begin{center}
        \epsfig{file=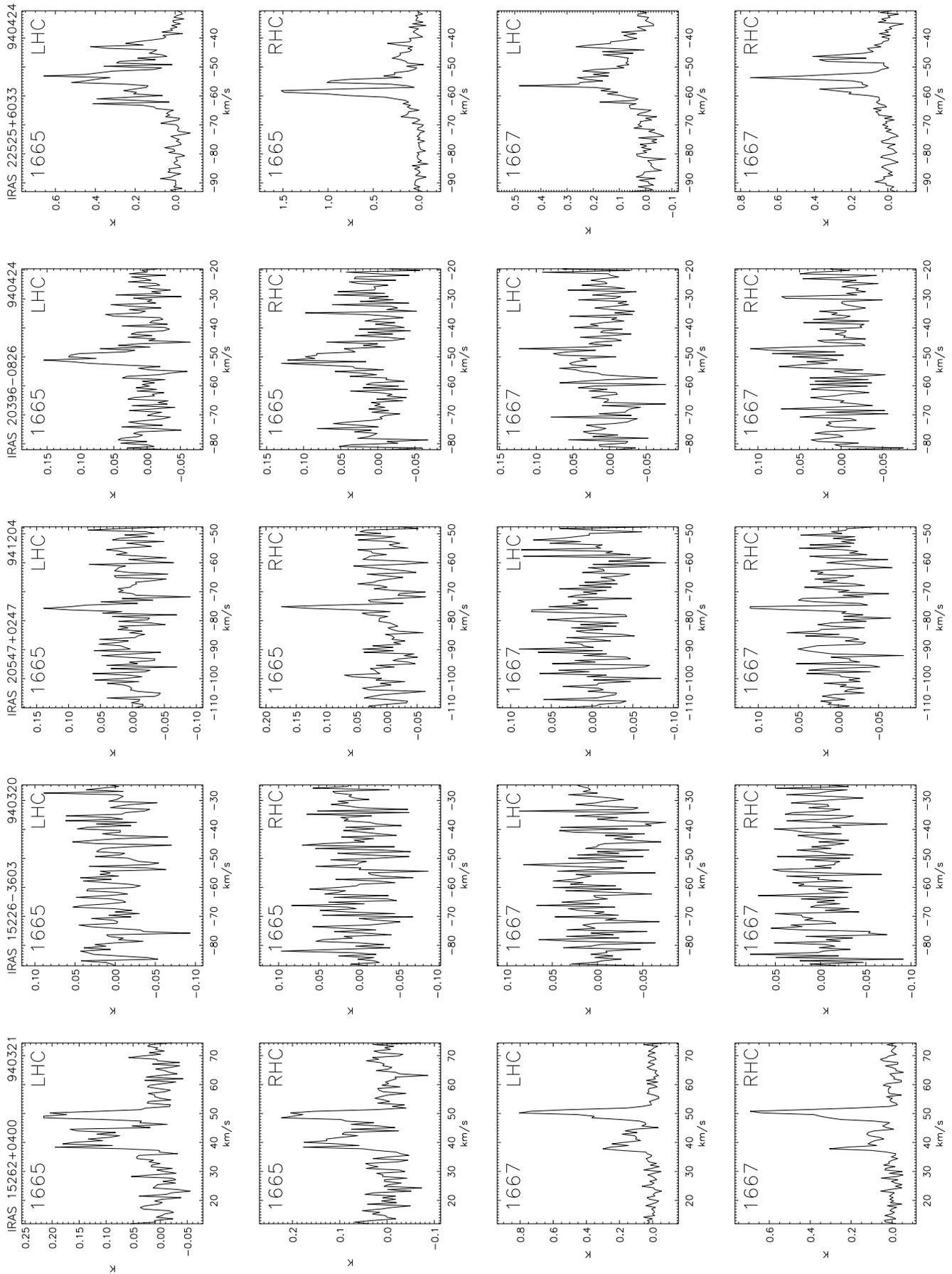,angle=-180}
   \end{center}
   \caption{Spectra of the sources at 1665 and 1667~MHz in the
	    left- (LHC) and right-handed (RHC) polarizations, appearing in 
	    increasing value of the [25$-$12] colour index.}
   \label{fig:profil spectral evolution}
\end{figure*}
\begin{figure*}
   \begin{center}
        \epsfig{file=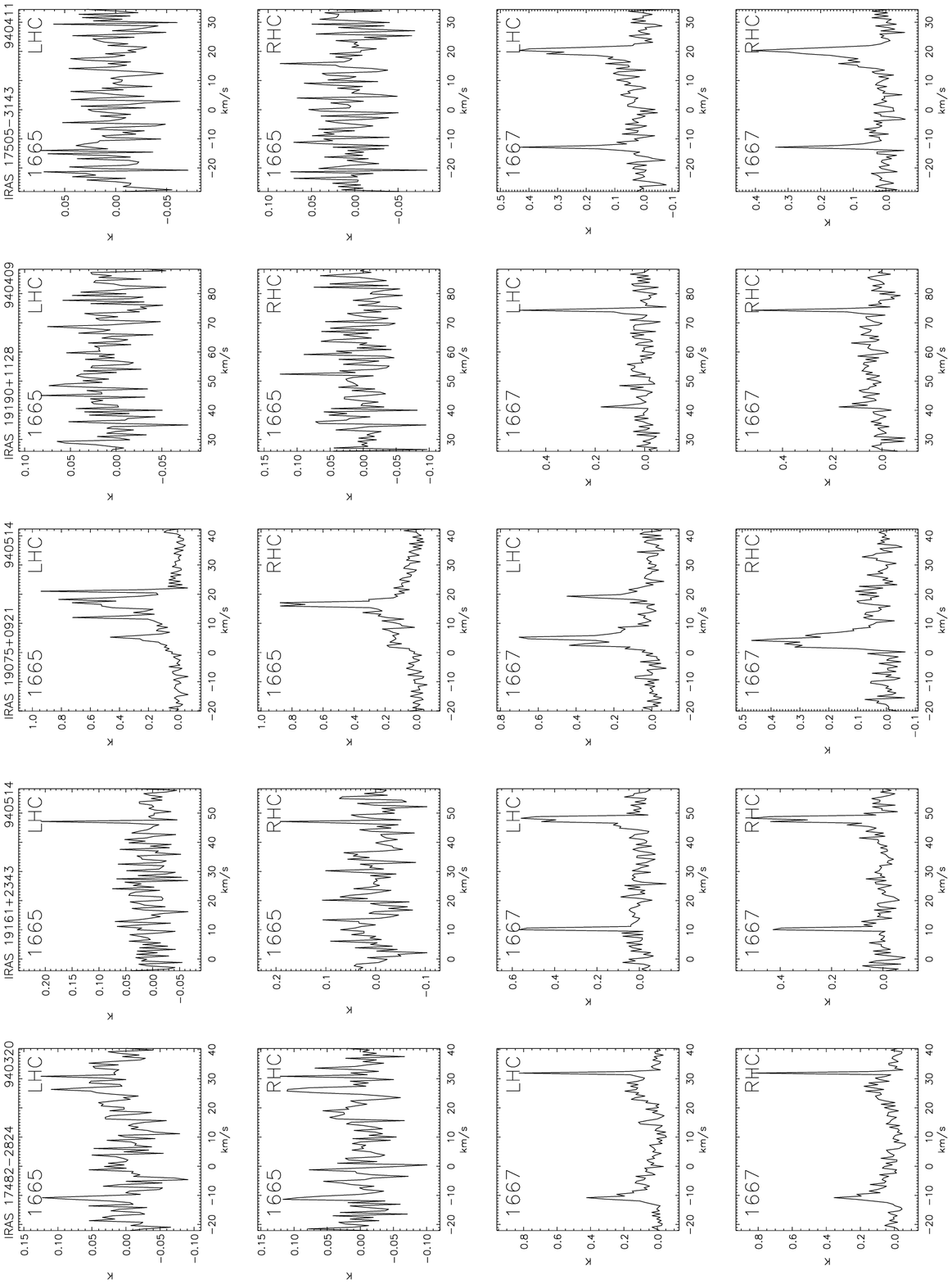,angle=-180}
   \end{center}
   \addtocounter{figure}{-1}
   \caption{\it cont.}
\end{figure*}
\begin{figure*}
   \begin{center}
        \epsfig{file=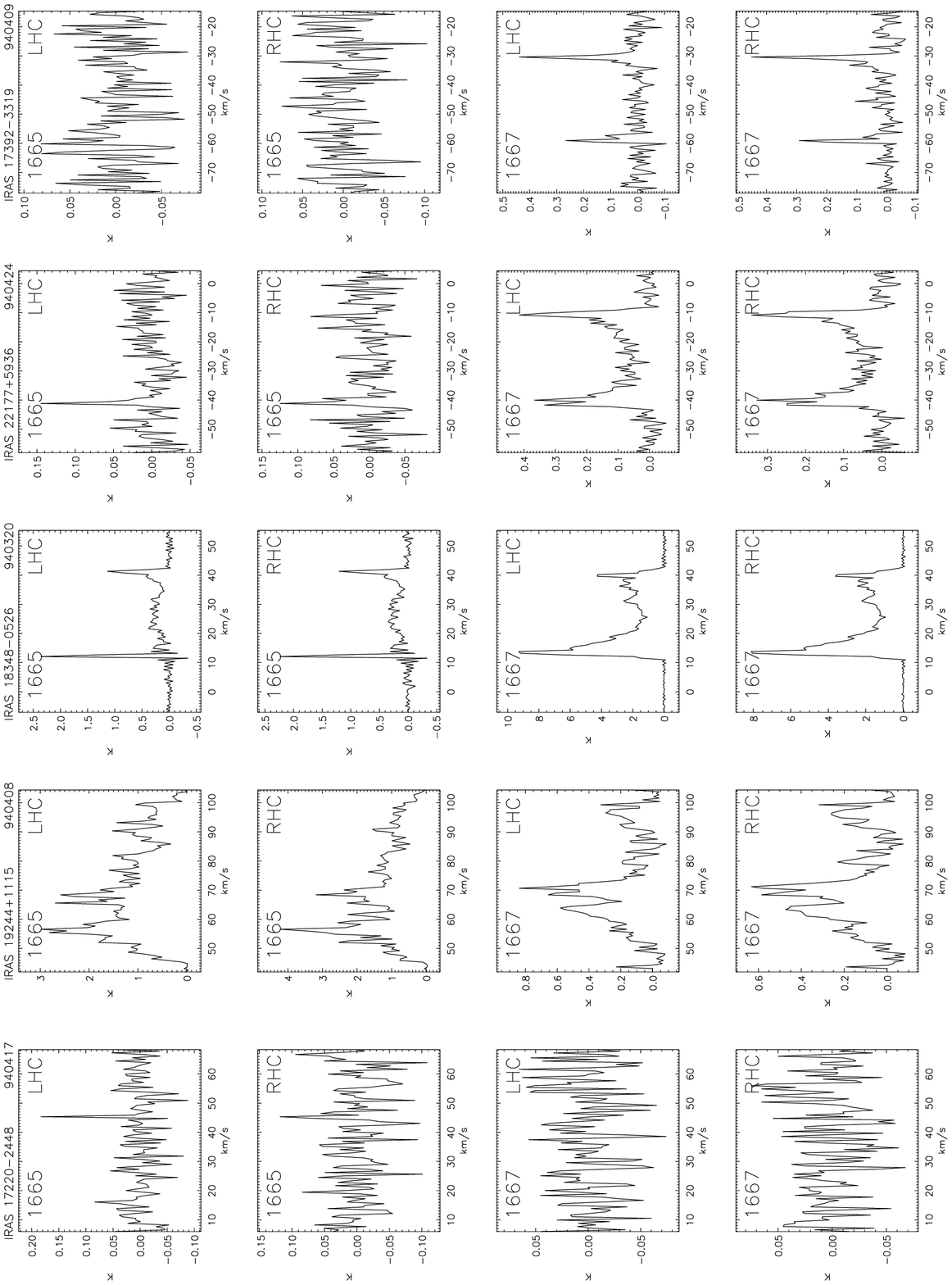,angle=-180}
   \end{center}
   \addtocounter{figure}{-1}
   \caption{\it cont.}
\end{figure*}
\begin{figure*}
   \begin{center}
        \epsfig{file=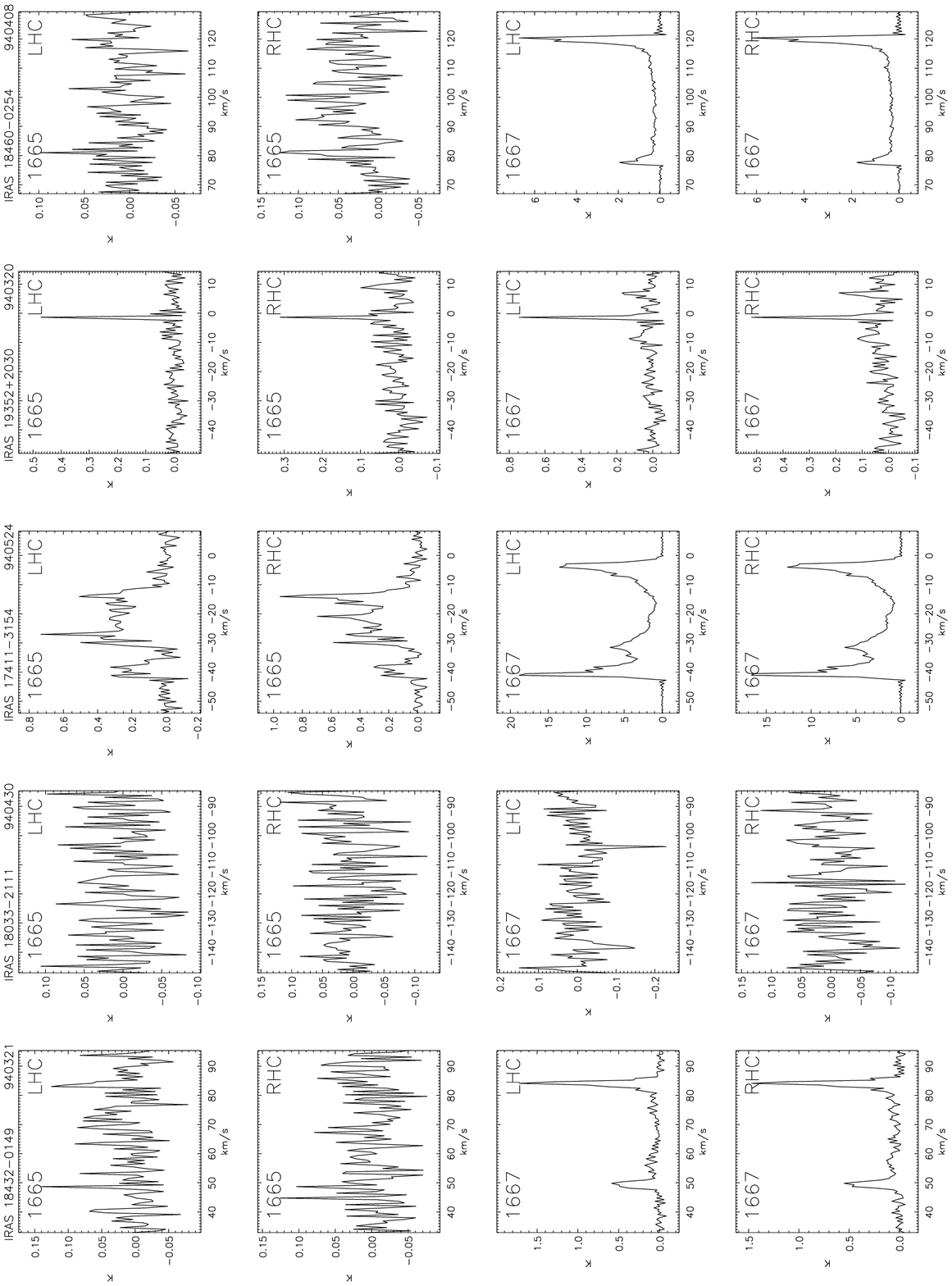,angle=-180}
   \end{center}
   \addtocounter{figure}{-1}
   \caption{\it cont.}
\end{figure*}
\begin{figure*}
   \begin{center}
        \epsfig{file=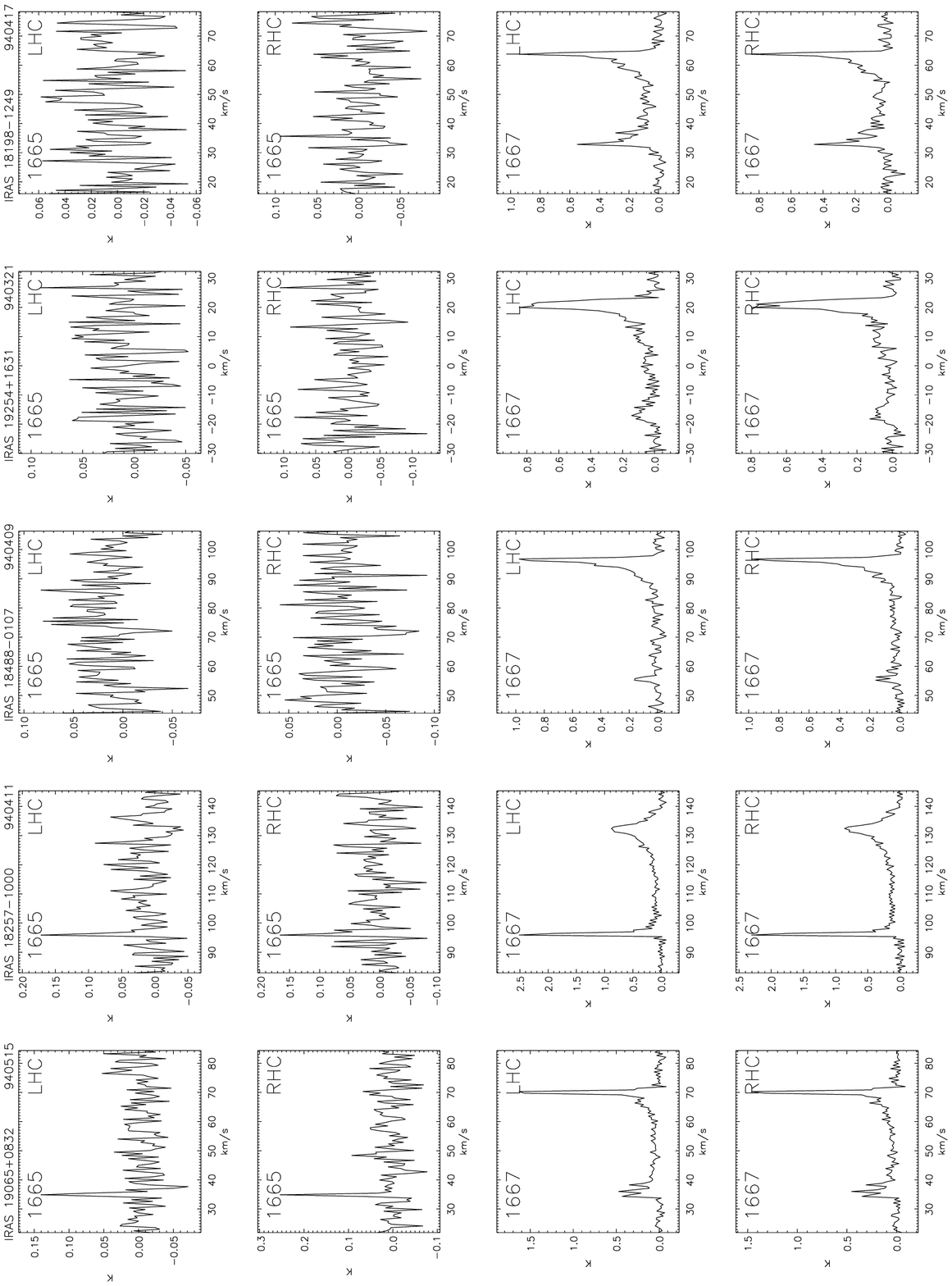,angle=-180}
   \end{center}
   \addtocounter{figure}{-1}
   \caption{\it cont.}
\end{figure*}
\begin{figure*}
   \begin{center}
        \epsfig{file=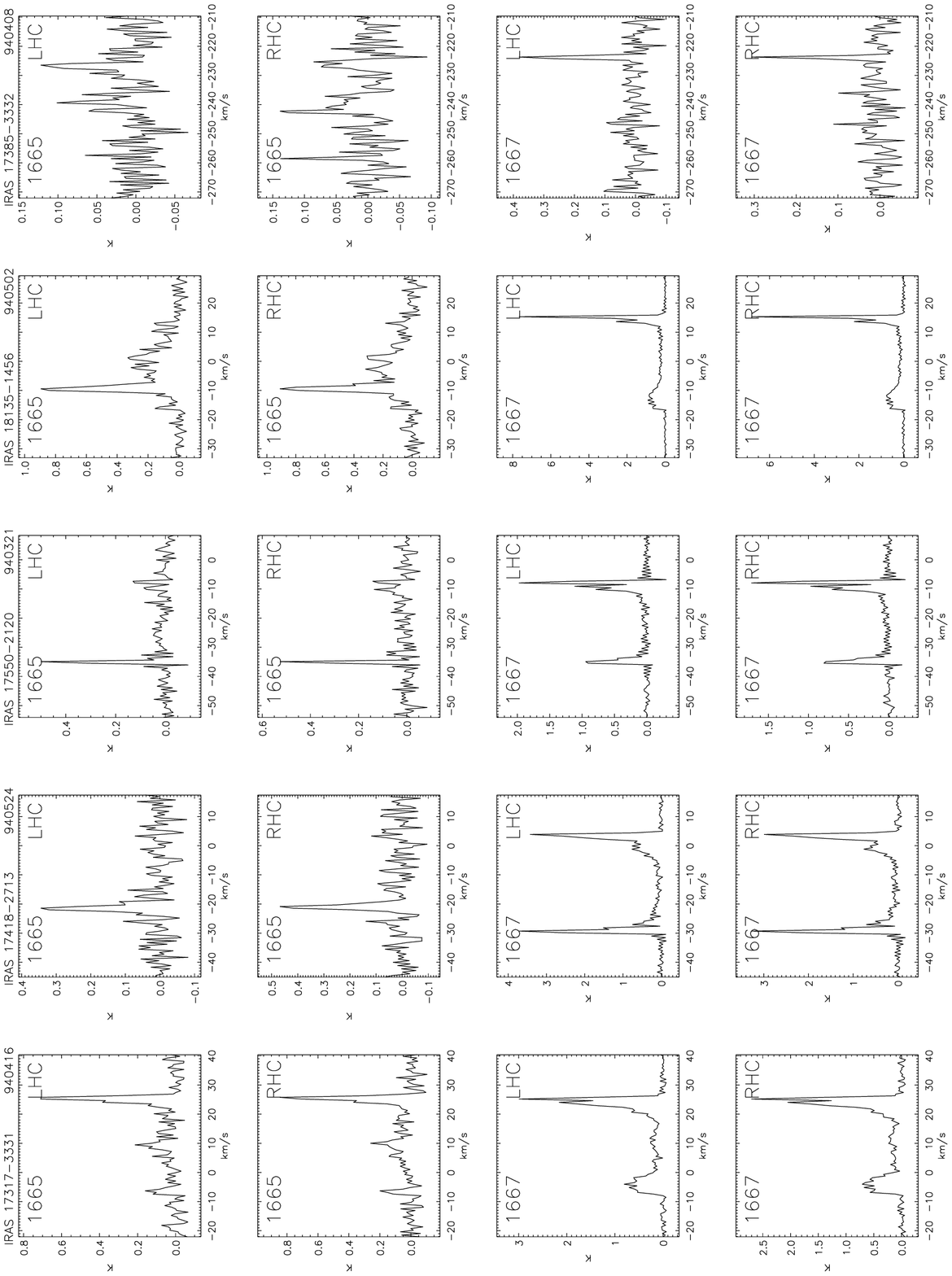,angle=-180}
   \end{center}
   \addtocounter{figure}{-1}
   \caption{\it cont.}
\end{figure*}
\section{Spectral measurement results} \label{appendix C}

\noindent
\begin{table*}
\caption{Results of the spectral measurements at 1665~MHz}
\label{Table:evolution spectrale 65}
		\scriptsize
\begin{tabular*}{16.75cm}{@{}@{\extracolsep{-0.188cm}}rlrrrrrrrrrrrrrr@{}}
\hline
 Iras Name & ps$^{2}$ & \multicolumn{2}{c}{V$_{b}^{1}$} & 
 \multicolumn{2}{c}{V$_{r}^{1}$} & \multicolumn{2}{c}{V$_{int}^{1,3}$} & 
 V$_{*}^{1}$ & V$_{exp}^{1}$ & 
 \multicolumn{2}{c}{S$_{b}^{1}$} & \multicolumn{2}{c}{S$_{r}^{1}$} & 
 \multicolumn{2}{c}{S$_{int}^{1,3}$} \\
            &                           &  \multicolumn{2}{c}{km/s}       & 
 \multicolumn{2}{c}{km/s}        & \multicolumn{2}{c}{km/s}            &  
 km/s        & km/s          & 
 \multicolumn{2}{c}{Jy}          & \multicolumn{2}{c}{Jy}          & 
 \multicolumn{2}{c}{Jy}             \\
            &                           &   L$^{1}$      &    R$^{1}$     &
       L        &        R       &       L        &        R       &
                &                &
       L        &        R       &       L        &        R       &
       L        &        R           \\
\hline
15226$-$3603 &  &     ... &     ... &     ... &     ... &         &         &
	                  &         &
	              ... &     ... &     ... &     ... &         &         \\
15262+0400 &  &   38.38 &   38.38 &   49.06 &   48.50 &         &         & 
                  43.6  &    5.2  &
		   0.19 &  0.18   &  0.22   &  0.22   &         &         \\
20396$-$0826 &c?p?&   ... &    ... &     ... &     ... &  $-$51.13 & $-$52.25 & 
                        &         &
	            ... &     ... &     ... &     ... &  0.16   &  0.13   \\
20547+0247 &  &     ... &     ... &$-$75.75 &$-$75.19 &          &         &
                        &         &
		    ... &     ... &  0.14   &  0.17   &          &        \\
22525+6033 &p &         &  $-$67.19?&      &         & $-$55.38 & $-$55.38 &  
			&           &
			&     0.19? &      &         &  0.56    &    1.01 \\
	& &	$-$62.69&           & $-$41.87 & $-$41.87 & $-$53.12 &$-$53.12&
                 $-$52.3 &     10.4 &
 		   0.41 &           &  0.25   &  0.34   &  0.66   &  0.38   \\
	& &	$-$61.00 &          & $-$43.00&         & $-$49.75 &        &
                         &          &
 		   0.39  &          &  0.43   &         &  0.36    &        \\
	& &	$-$58.19 & $-$58.19 &         & $-$46.94& $-$48.62 &        &
                         &          &
 		   0.26  &   1.52   &         &    0.29 &  0.29    &        \\
\hline
17482$-$2824 &  & 10.81 &  $-$11.38 &   30.81 &   30.81 &         &         &
                   9.9  &      20.9 &
 		  0.12  &  0.12   &  0.13   &  0.12   &         &         \\
17505$-$3143 &  &   ... &  ... &     ... &     ... &         &         &
                        &         &
  		    ... &  ...  &     ... &     ... &         &         \\

19075+0921 &p &    5.25 &    2.44 &   21.00 &   17.06 &   12.00 &    8.06 &
                  11.4  &    7.6  &
		   0.46 &  0.19   &  0.94   &  0.87   &  0.72   &  0.22   \\
           &  &     ... &     ... &     ... &     ... &   17.06 &   10.88 &
                        &         &
 		    ... &     ... &     ... &     ... &  0.73   &  0.22   \\
           &  &     ... &     ... &     ... &     ... &   18.19 &   13.69 & 
                        &         &
 		    ... &     ... &     ... &     ... &  0.82   & 0.34    \\
19161+2343 &  &     ... &     ... &   47.13 &   47.13 &         &         &
                        &         &
 		    ... &     ... &  0.21   &  0.19   &         &         \\
19190+1128 &  &     ... &    ...  &     ... &     ... &         &         &
                        &         &
 		    ... &   ...   &     ... &     ... &         &         \\
19244+1115 &p &   56.56 &   56.56 &   99.31 &   98.13 &   65.56 &   58.81 &
                  77.6  &   21.1  &
		  2.99  & 4.23    & 1.05    & 0.97    & 2.69    & 2.56  \\
           &  &     ... &     ... &     ... &     ... &   68.38 &   61.62 &
                        &         &
		    ... &     ... &     ... &     ... &  2.59   &  2.27   \\   
           &  &     ... &     ... &     ... &     ... &   76.81 &   65.56 &
                        &         &
		    ... &     ... &     ... &     ... &  1.60   &  2.29   \\
           &  &     ... &     ... &     ... &     ... &   81.88 &   68.38 &
                        &         &
		    ... &     ... &     ... &     ... &  1.51   &  3.19   \\
           &  &     ... &     ... &     ... &     ... &   90.31 &   76.25 &
                        &         &
		    ... &     ... &     ... &     ... &  1.53   &  1.68   \\
           &  &     ... &     ... &     ... &     ... &   93.13 &   81.88 &
                        &         &
		    ... &     ... &     ... &     ... &  1.43   &  1.32   \\
           &  &     ... &     ... &     ... &     ... &         &   90.88 &
                        &         &
		    ... &     ... &     ... &     ... &         &  1.56   \\
           &  &     ... &     ... &     ... &     ... &         &   93.13 &
                        &         &
		    ... &     ... &     ... &     ... &         &  1.06   \\
           &  &     ... &     ... &     ... &     ... &         &   96.50 &
                        &         &
		    ... &     ... &     ... &     ... &         &  0.95   \\
\hline
17317$-$3331 &ci& $-$6.31 &  $-$6.31 &   25.19 &   25.75 &    9.44 &   10.00 & 
                     9.6  &    15.9  &
		     0.16 &    0.20 &    0.71 &    0.85 &    0.21 &  0.26   \\
17220$-$2448 &  &   ... &     ... &   45.31 &   45.31 &         &         & 
                        &         &
 		    ... &     ... &    0.18 &    0.12 &         &         \\
17385$-$3332 &ci &$-$ ... & $-$258.25 & $-$226.50 & $-$225.37 & $-$239.44 & $-$242.15 & 
                   $-$241.8  &    16.2 &
		   0.06?&    0.14 &    0.12 &    0.09 &    0.10 &  0.14   \\
17392$-$3319 &  & $-$60.19?&   ... &  $-$22.50?&     ... &         &         & 
                   $-$41.3?&  18.8?&
		   0.08?&     ... &    0.07?&     ... &         &         \\
17411$-$3154 &p & $-$38.31 & $-$38.31 & $-$14.13 & $-$14.13 & $-$27.06 & $-$20.88 & 
                   $-$26.2 &   12.1 &
		   0.32 &    0.59 &    0.51 &    0.96 &    0.73 &  0.59   \\
17418$-$2713 &ci &    ... &     ... &   ... &     ... &$-$21.44 & $-$20.88 &   
                          &         &
		      ... &     ... &   ... &     ... & 0.34   &    0.47 \\
17550$-$2120 &  & $-$34.94 & $-$34.94 & $-$7.37 & $-$7.37 &         &         & 
                   $-$21.1 &   13.8   &
		   0.50 &    0.53 &    0.13 &    0.14 &         &         \\
18033$-$2111 &  &   ... &     ... &     ... &     ... &         &         & 
                        &         &
		    ... &     ... &     ... &     ... &         &         \\
18135$-$1456 &p & $-$16.19 & $-$16.19 & 13.06 &  13.06 &   $-$9.44 & $-$9.44 & 
                    $-$1.6 &     14.6 &
		   0.16 &    0.15 &    0.16 &    0.18 &    0.90 &  0.91   \\
           &  &     ... &     ... &     ... &     ... &   $-$2.13 &   $-$2.67 & 
                        &         &
		    ... &     ... &     ... &     ... &    0.31 &  0.32   \\
           &  &     ... &     ... &     ... &     ... &    1.25 &    1.81 & 
                        &         &
		    ... &     ... &     ... &     ... &    0.33 & 0.31    \\
           &  &     ... &     ... &     ... &     ... &    5.75 &    5.75 & 
                        &         &
		    ... &     ... &     ... &     ... &    0.16 &  0.14   \\
           &  &     ... &     ... &     ... &     ... &   10.25 &         & 
                        &         &
		    ... &     ... &     ... &     ... &    0.18 &         \\
18198$-$1249 &  &   ... &     ... &     ... &     ... &         &         & 
                        &         &
		    ... &     ... &     ... &     ... &         &         \\
18257$-$1000 &  &   95.87 &   95.87 &    ... &     ... &         &         & 
                          &         &
		     0.17 &    0.17 &    ... &     ... &         &         \\
18348$-$0526 &s &   12.06 &   12.06 &   41.31 &   41.31 &   26.69 &   26.13 & 
                    26.7  &   14.6  &
		   2.36 &    2.24 &    1.14 &    1.21 &    0.31 &  0.30   \\
18432$-$0149 &  &   48.69 &     ... &   83.00 &     ... &         &         & 
                    65.8  &    17.2 &
		   0.14 &     ... &    0.13 &     ... &         &         \\
18460$-$0254 &s?&   81.00 &   81.00 &     ... &     ... &         &  100.69?& 
                        &         &
		   0.10 &    0.12 &     ... &     ... &         &  0.12?  \\
18488$-$0107 &  &     ... &     ... &     ... &     ... &         &         & 
                        &         &
		    ... &     ... &     ... &     ... &         &         \\
19065+0832 &  &   34.87 &   34.87 &     ... &     ... &         &         & 
                        &         &
		   0.14 &    0.25 &     ... &     ... &         &         \\
19254+1631 &  &     ... &     ... &     ... &     ... &         &         & 
                        &         &
		    ... &     ... &     ... &     ... &         &         \\
19352+2030 &ci&     ... &     ... &     ... &    8.75?&   $-$1.37 &   $-$1.37 & 
                        &         &
		    ... &     ... &     ... &    0.10?&    0.47 &    0.31 \\
22177+5936 &  & $-$41.19 & $-$41.19 &   ... & $-$11.37?&         &         & 
                $-$26.3? &   14.9? &
		  0.14   &  0.13   &     ... &  0.08?  &         &         \\
\hline
\end{tabular*} \\
			\normalsize
\small{
(1)  {\bf L}: left-handed polarization, 
     {\bf R}: right-handed polarization \\
(2)  {\bf profile shape characteristic} : \\
     {\bf p}: plateau profile,
     {\bf c}: profile with only a single peak centred on the stellar velocity,
     {\bf s}: ``standard'' profile with inter-peak signal,
     {\bf ci}: profile with an internal component or group of components  
         (well detached from the two standard peaks) \\
(3)  {\bf V$_{int}$}: in one of the mentioned cases. When the profile is 
	   triple-peaked, the velocity of the component showing
	   the strongest intensity (the latter might not be centred on the 
	   stellar velocity) is at least given.
     {\bf S$_{int}$}: intensity corresponding to V$_{int}$ \\
}
\end{table*}

\noindent
\begin{table*}

\caption{Results of the spectral measurements at 1667~MHz} 

\label{Table:evolution spectrale 67}
		\scriptsize
\begin{tabular*}{16.75cm}{@{}@{\extracolsep{-0.188cm}}rlrrrrrrrrrrrrrr@{}}
\hline
 Iras NAme  & ps$^{2}$ & \multicolumn{2}{c}{V$_{b}^{1}$} & 
 \multicolumn{2}{c}{V$_{r}^{1}$} & \multicolumn{2}{c}{V$_{int}^{1,3}$} & 
 V$_{*}^{1}$ & V$_{exp}^{1}$ & 
 \multicolumn{2}{c}{S$_{b}^{1}$} & \multicolumn{2}{c}{S$_{r}^{1}$} & 
 \multicolumn{2}{c}{S$_{int}^{1,3}$} \\
            &                           &  \multicolumn{2}{c}{km/s}       & 
 \multicolumn{2}{c}{km/s}        & \multicolumn{2}{c}{km/s}            &  
 km/s        & km/s          & 
 \multicolumn{2}{c}{Jy}          & \multicolumn{2}{c}{Jy}          & 
 \multicolumn{2}{c}{Jy}             \\
            &                           &   L$^{1}$      &    R$^{1}$     &
       L        &        R       &       L        &        R       &
                &                &
       L        &        R       &       L        &        R       &
       L        &        R           \\
\hline
15226$-$3603 &  &     ... &     ... &     ... &     ... &         &         &
                          &         &
		      ... &     ... &     ... &     ... &         &         \\
15262+0400 &  &   37.82 &   37.82 &   50.18 &   50.74 &         &         &
                  44.1  &    6.3  &
		   0.30 &    0.31 &    0.81 &    0.69 &         &         \\
20396$-$0826 & &    ... & $-$53.37?& $-$47.19 &  $-$47.19 &      &         &
                   50.3?&   3.1?  &
		        & 0.07?    &    0.12  &    0.11   &       &        \\
20547+0247 &  &     ... &     ... &     ... & $-$75.19 &         &          &
                        &         &
		    ... &     ... &     ... &   0.11   &         &         \\
22525+6033 &  & $-$62.12 & $-$57.63 & $-$43.02 & $-$46.39 &         &         &
                 $-$52.6 &     9.5 &
		   0.17 &    0.37 &    0.27 &    0.41 &         &         \\
           &  & $-$56.51 & $-$53.70 & $-$45.83 & $-$47.51 &         &         &
                        &         &
		   0.48 &    0.75 &    0.16 &    0.37 &         &         \\
           &  &  $-$53.70 &     ... &  $-$44.71 &     ... &         &         &
                        &         &
		   0.26 &     ... &    0.17 &     ... &         &         \\
           &  &  $-$51.45 &     ... &     ... &     ... &         &         &
                        &         &
		   0.24 &     ... &     ... &     ... &         &         \\
\hline
17482$-$2824 &ci& $-$10.79 & $-$10.79 &  31.91 &   31.91 &   15.06 &         &
                     10.6  &    21.3  &
		   0.43 &    0.35 &    0.83 &    0.84 &    0.11 &         \\
17505$-$3143 &  & $-$12.86 & $-$12.86 &  20.29 &   20.29 &         &         &
                      3.7  &    16.6  &
		   0.38 &    0.33 &    0.48 &    0.46 &         &         \\
19075+0921 &  &    5.26 &    4.13 &   19.30 &   20.99 &         &         &
                  12.4  &     7.7 &
		   0.70 &    0.47 &    0.45 &    0.11 &         &         \\
19161+2343 &  &   10.02 &   10.02 &   48.23 &   48.23 &         &         &
                  29.1  &   19.1  &
		   0.57 &    0.43 &    0.56 &    0.51 &         &         \\
19190+1128 &  &   41.14 &   41.14 &   74.29 &   74.29 &         &         &
                  57.7  &   16.6  &
		   0.18 &    0.17 &    0.51 &    0.52 &         &         \\
19244+1115 &ci&   63.89 &   63.32 &   99.29 &   99.29 &         &         &
                  81.4  &   17.8  &
		   0.58 &    0.47 &    0.33 &    0.32 &         &         \\
           &  &   70.63 &   71.19 &     ... &     ... &         &   79.62 &
                        &         &
		   0.84 &    0.63 &     ... &     ... &         &    0.23 \\ 
\hline
17220$-$2448 &  &     ... &     ... &     ... &     ... &         &         &
                        &         &
		    ... &     ... &     ... &     ... &         &         \\
17317$-$3331 &s & $-$4.05 & $-$4.05 &  25.17 &   25.17 &   10.00 &   10.00 &
                  $-$10.6 &   14.6  &
		   0.81 &    0.71 &    2.99 &    2.71 &    0.26 &    0.23 \\
17385$-$3332 &ci?&  ... &     ... & $-$223.71 & $-$223.71 & $-$246.18?& $-$246.74?&
                        &         &
		    ... &     ... &    0.38 &    0.31 &    0.09?&    0.11?\\
           &  &     ... &     ... &     ... &     ... &         & $-$236.07?&
                        &         &
		    ... &     ... &     ... &     ... &         &    0.10?\\
17392$-$3319 &  & $-$59.05 & $-$59.05 & $-$30.39 & $-$30.39 &       &         &
                   $-$44.7 &     14.3 &
		   0.27 &    0.29 &    0.44 &    0.45 &         &         \\
17411$-$3154 &s & $-$41.10 & $-$41.10 & $-$4.02 & $-$4.02 & $-$22.66 & $-$22.56 &
                   $-$22.6 &    18.5  &
		  18.87 &   16.67 &   13.55 &   12.67 &    1.93 &    1.69 \\
17418$-$2713 &  & $-$29.29 & $-$29.29 &  3.86 &    3.86 &         &         &
                   $-$12.7 &    16.6  &
		   3.72 &    3.28 &    3.43 &    2.99 &         &         \\
17550$-$2120 &  & $-$34.92 & $-$35.49 & $-$7.95 & $-$7.95 &         &         &
                  $-$21.6  &     13.6 &
		   0.94 &    0.81 &    1.98 &    1.71 &         &         \\
18033$-$2111 &  &     ... &     ... &     ... &     ... &         &         &
                        &         &
		    ... &     ... &     ... &     ... &         &         \\
18135$-$1456 &s & $-$13.36 & $-$13.36 &  15.29 &   15.29 &    1.24 &    1.24 &
                       1.0 &    14.3  &
		   0.91 &    0.79 &    7.67 &    6.82 &    0.30 &    0.15 \\
18198$-$1249 &s &   32.83 &   32.83 &   63.73 &   63.73 &   48.56 &   48.56 &
                     48.3 &    15.4 &
		   0.55 &    0.46 &    0.94 &    0.85 &     0.12 &     0.07 \\
18257$-$1000 &s &   95.90 &   95.90 &  132.42 &  132.42 &  114.44 &   114.44 &
                    114.2 &   18.3  &
		   2.52 &    2.31 &    0.88 &    0.86 &     0.11  &    0.13 \\
18348$-$0526 &s&   13.76 &   13.76 &   39.61 &   39.61 &   26.12 &   26.12 &
                     26.7 &   12.9  &
		   9.27 &    8.11 &    4.26 &    3.60 &    1.41 &    1.26 \\
18432$-$0149 &  &   49.83 &   49.83 &   84.10 &   84.10 &         &         &
                    67.0  &    17.1 &
		   0.59 &    0.55 &    1.72 &    1.47 &         &         \\
18460$-$0254 &s &   77.65 &   77.65 &  120.35 &  120.35 &   99.00 &   99.00 &
                     99.0 &    21.3 &
		   1.94 &    1.78 &    6.76 &    6.22 &    0.25 &    0.30 \\
18488$-$0107 &  &   55.21 &   55.21 &   96.79 &   96.79 &         &         &
                    76.0  &   20.8 &
		   0.16 &    0.16 &    0.98 &    0.98 &         &         \\
19065+0832 &s?&   36.02 &   36.02 &   70.29 &   70.29 &   53.44? &   54.07? &
                   53.2 &   17.1  &
		   0.47 &    0.46 &    1.63 &    1.46 &    0.09? &    0.07? \\
19254+1631 &  &  $-$17.10 &  $-$18.23 &   19.98 &   19.98 &         &         &
                     1.2  &     18.8  &
		   0.14 &    0.13 &    0.85 &    0.80 &         &         \\
19352+2030 &ci& $-$8.70 & $-$8.70 &   7.04 &    7.04 &   $-$1.39 &   $-$1.39 &
                 $-$0.8 &    7.9  &
		   0.13 &    0.12 &    0.17 &    0.19 &    0.75 &    0.52 \\
22177+5936 &s & $-$40.05 & $-$40.05 & $-$10.83 & $-$10.83 & $-$26.00 &  26.56 &
                 $-$25.4 &    14.6  &
		   0.37 &    0.33 &    0.42 &    0.34 &    0.06 &    0.06 \\
\hline
\end{tabular*} \\
			\normalsize
\small{
1,2 and 3~: same as for the previous table
}
\end{table*}


\end{document}